\documentclass[JHEP,12pt]{article}
\usepackage{jheppub}
\usepackage{amssymb,amsmath,amsthm}
\usepackage[shortlabels]{enumitem}
\usepackage{braket}
\usepackage{comment} 
\usepackage{subcaption}
\usepackage{graphicx, xcolor, varwidth}
\usepackage{color}
\newtheorem{thm}{Theorem}
\newtheorem{cor}[thm]{Corollary} 

\theoremstyle{remark}
\newtheorem{defn}[thm]{Definition}
\newtheorem{rem}[thm]{Remark}
\newtheorem{conj}[thm]{Conjecture}
\newtheorem{conv}[thm]{Convention}

\newcommand{\deltabar}{\delta\hspace*{-0.2em}\bar{}\hspace*{0.2em}}

\DeclareMathOperator{\cl }{cl}

\DeclareMathOperator{\A}{Area}

\DeclareMathOperator{\minEW}{minEW}
\DeclareMathOperator{\maxEW}{maxEW}
\DeclareMathOperator{\EW}{EW}
\renewcommand{\S}{S_{\rm gen}}
\newcommand{\emax}{e_{\rm max}}
\newcommand{\emin}{e_{\rm min}}
\newcommand{\smax}{\Sigma_{\rm max}}
\newcommand{\smin}{\Sigma_{\rm min}'}
\newcommand{\ba}{\mathbf{a}}
\newcommand{\bb}{\mathbf{b}}

\author{Raphael Bousso and Geoff Penington}

\affiliation{Center for Theoretical Physics and Department of Physics,\\
University of California, Berkeley, California 94720, U.S.A. 
} 

\emailAdd{bousso@berkeley.edu}
\emailAdd{geoffp@berkeley.edu}

\title{Holograms In Our World}

\abstract{In AdS/CFT, the entanglement wedge $\EW(B)$ 
is the portion of the bulk geometry that can be reconstructed from a boundary region $B$; in other words, $\EW(B)$ is the hologram of $B$. We extend this notion to arbitrary spacetimes
. Given any gravitating region $a$, we define a max- and a min-entanglement wedge, $\emax(a)$ and $\emin(a)$, such that $\emin(a)\supset \emax(a)\supset a$.

Unlike their analogues in AdS/CFT, these two spacetime regions can differ already at the classical level, when the generalized entropy is approximated by the area. All information outside $a$ in $\emax(a)$ can flow inwards towards $a$, through quantum channels whose capacity is controlled by the areas of intermediate homology surfaces. In contrast, all information outside $\emin(a)$ can flow outwards.

The generalized entropies of appropriate entanglement wedges obey strong subadditivity, suggesting that they represent the von Neumann entropies of ordinary quantum systems. The entanglement wedges of suitably independent regions satisfy a no-cloning relation. This suggests that it may be possible for an observer in $a$ to summon information from spacelike related points in $\emax(a)$, using resources that transcend the semiclassical description of $a$.}

\makeatletter
\gdef\@fpheader{\mbox{}}
\makeatother

\begin{document}
\maketitle

\section{Introduction}
\paragraph{Background}

Entanglement wedges have been at the core of much of the recent progress in our understanding of quantum gravity. In AdS/CFT, an entanglement wedge $\EW(B)$ is a gravitating (or ``bulk'') spacetime region that is holographically dual to a given spatial region $B$ (at some fixed time) on the boundary \cite{Bousso:2012sj,Czech:2012bh,Wall:2012uf,Headrick:2014cta,Jafferis:2015del,
Dong:2016eik,Cotler:2017erl}. More precisely, $\EW(B)$, if it exists, satisfies the following two properties:
\begin{enumerate}[a.]
    \item A sufficiently simple quasi-local bulk operator in $\EW(B)$ can be implemented by a CFT operator in the algebra associated to the region $B$.
    \item No bulk operator outside $\EW(B)$ can be so implemented.
\end{enumerate}
In the language of the holographic principle~\cite{tHooft:1993dmi,Susskind:1994vu,Fischler:1998st,Bousso:1999cb}, $\EW(B)$ captures the depth of the hologram that pops out from $B$.\footnote{In fact, one of the authors feels that entanglement wedges should simply be renamed ``holograms'' --- starting with this paper. After many hours of heated debate, we will continue to use ``entanglement wedge.'' For now.} 

Given the entire spacetime $M$ dual to a particular boundary state, there is a standard prescription for finding the entanglement wedge $\EW(B)$ of a boundary subregion $B$, as follows. The generalized entropy of a partial Cauchy slice $c$ in a gravitating spacetime $M$ with Newton's constant $G$ is
\begin{align} \label{eq:sgen}
    \S(c) = \frac{\mathrm{Area}(\partial c)}{4G\hbar} + S(c)~,
\end{align}
where $S$ denotes the von Neumann entropy of the bulk quantum state reduced to $c$, and $\partial c$ denotes the boundary of $c$ in a full Cauchy slice of $M$. $\EW(B)$ is defined as the domain of dependence of a spatial bulk region $c$ whose only asymptotic boundary is $B$, and which is quantum extremal, meaning that the generalized entropy $\S(c)$ is stationary under small deformations $\partial c$ of the boundary of $c$ inside $M$.  If there is more than one such domain of dependence, $\EW(B)$ is the one with smallest $\S$. 

The entanglement wedge can be used to compute the entropy $S(\rho_B)$ of the reduced boundary state $\rho_B$ on the region $B$ as \cite{Ryu:2006bv,Hubeny:2007xt,Faulkner:2013ana, Engelhardt:2014gca}
\begin{equation}\label{eq-rt}
    S(\rho_B)= \S[\EW(B)]~.
\end{equation}
Notably, this formula can be justified (with some assumptions) by applying the gravitational path integral to the computation of Renyi entropies~\cite{Lewkowycz:2013nqa}, without appealing to string theory or other microscopic details of the theory. 

When external quantum systems are coupled to the spacetime, the definition of an entanglement wedge needs to be generalized.
Consider for example a black hole in anti-de Sitter space evaporating into a nongravitational bath $R$, and let $B(t)$ be the entire boundary at the time $t$. Then $\EW[B(t)]$ has a phase transition at the Page time~\cite{Penington:2019npb,Almheiri:2019psf}. After this time, most of the interior of the black hole is not contained in the entanglement wedge of the black hole's asymptotic boundary. 

If the information inside the black hole is not encoded in $B$, then where did it go? One can formally define an entanglement wedge $\EW(R)$ for the external bath, by augmenting the standard prescription to specify that $R$ itself must always be included in $\EW(R)$~\cite{Hayden:2018khn, Penington:2019npb, Almheiri:2019hni}.\footnote{This augmented prescription can be derived using the same gravitational replica trick techniques as the standard prescription~\cite{Penington:2019kki, Almheiri:2019qdq}.} Before the Page time, one finds $\EW[R(t)] = R(t)$. But after the Page time, $\EW[R(t)]$ also contains a portion of the black hole interior. This disconnected ``island'' is precisely the spacetime region that was missing from $\EW[B(t)]$.  Its inclusion yields the Page curve~\cite{Page:1993df, Page:1993wv} for the entropy of Hawking radiation, and it implies the Hayden-Preskill criterion~\cite{Hayden:2007cs} for the information that can be recovered from the radiation at the time $t$.

The fact that an external system $R$ can acquire a nontrivial entanglement wedge within a gravitating spacetime was a radical new development. It means that the concept of an entanglement wedge can be divorced from the conformal boundary of an asymptotically-AdS spacetime.

But if Hawking radiation can have an entanglement wedge, then it should possess this property even before being extracted from the spacetime. Our universe, for example, is gravitating, and it does not appear to have an asymptotically-AdS boundary. Yet one still obtains the Page curve by associating an entanglement island to the Hawking radiation of a black hole~\cite{Bousso:2020kmy,Dong:2020uxp}. It is natural then to ask for the most general class of objects to which one can associate an entanglement wedge.

This motivated us to propose a significant generalization of the notion of an entanglement wedge~\cite{Bousso:2022hlz}: we conjectured that any gravitating region $a$ has an associated \emph{generalized entanglement wedge}. For example, $a$ can be inside a black hole, or part of a closed universe.
We conjectured, moreover, that no separate rule is needed to associate an entanglement wedge to a portion $B$ of the conformal boundary of AdS, or to an auxiliary nongravitational system $R$. The traditional entanglement wedges $\EW(B)$, $\EW(R)$ should arise as limiting cases of the generalized entanglement wedge, when the input bulk region $a$ is an asymptotic region with boundary $B$, or when the gravitational coupling $G$ is taken to zero, respectively.

In time-reflection symmetric spacetimes, we found  a simple proposal that meets these criteria: the generalized entanglement wedge of the region $a$ has the smallest generalized entropy among regions that contain $a$. Moreover, this proposal satisfies nontrivial properties expected of an entanglement wedge, such as nesting, no-cloning, and strong subadditivity~\cite{Bousso:2022hlz}. However, we did not succeed in formulating a proposal for general, time-dependent settings that satisfied all of these properties.

\paragraph{Max- vs.\ Min-Entanglement Wedges and State Merging}

In order to overcome this difficulty, it will be vital to absorb a seemingly unrelated development in our understanding of entanglement wedges: for generic bulk quantum states, $EW(B)$ may not exist, because
no region simultaneously satisfies both properties (a) and (b) above~\cite{Akers:2020pmf}. However, it is possible to define two bulk regions, $\maxEW(B) \subset \minEW(B)$, which are optimal with respect to each criterion separately~\cite{Akers:2020pmf,Akers:2023fqr}.

The max-entanglement wedge --- so named because its definition invokes the smooth conditional max-entropy~\cite{RenWol04a,Konig_2009} --- is the largest possible bulk region within which any quasi-local bulk operator can be reconstructed from the boundary region $B$. In contrast, the min-entanglement wedge is the smallest bulk region outside which no operator is reconstructible from $B$. Its definition involves the smooth conditional min-entropy. In general, $\minEW(B)$ may be strictly larger than $\maxEW(B)$, so that no single $EW(B)$ exists.

The smooth conditional max- and min-entropies are modifications of the usual conditional von Neumann entropy, developed in the study of one-shot quantum Shannon theory. Consider the communication task of \emph{quantum state merging} \cite{horodecki2007quantum}. The goal of this task is to obtain the state of a system $c$ with access only to a subsystem (or subregion) $\tilde c$ along with a minimal number of additional qubits. 

When merging a large number of copies of $c$, the number of qubits required, per copy, is quantified by the conditional von Neumann entropy $S(c)-S(\tilde c)$. Note that the conditional entropy need not be positive. For example, Bell pairs shared by $c\setminus \tilde c$ and $\tilde c$ give a negative contribution, because they can be used to teleport information into $\tilde c$.\footnote{It is important here that, along with the minimum qubits, one can also send unlimited free classical bits \cite{horodecki2007quantum} (or, more generally, zero-bits \cite{hayden2020approximate,Hayden:2018khn,Akers:2020pmf}).} This helps minimize the number of qubits that need to be sent. Thus, state merging can be accomplished with no additional qubits if $S(c)-S(\tilde c)\leq 0$. When the systems in question are geometric, each region's boundary contributes $\A/(4G\hbar)$ to the entropy, so this condition becomes $\S(c)-\S(\tilde c)\leq 0$.

When only a single copy is present, the number required is instead controlled by the conditional max-entropy $H_{\rm max}^\epsilon
$~\cite{berta2009single}. (The conditional min-entropy appears in closely related one-shot communication tasks.) Unlike von Neumann entropies, conditional max- and min-entropies cannot be written as a difference of entropies; consequently they are somewhat harder to work with. Again adding the area terms for geometric regions, one-shot quantum state merging can be accomplished with no additional qubits if $\A(c)/(4G\hbar)-\A(\tilde c)/(4G\hbar) +H_{\rm max}^\epsilon
\leq 0$.

For sufficiently nice bulk quantum states, conditional min- and max-entropies are equal; in such situations they can be replaced by the simpler conditional von Neumann entropy. For convenience, we will assume that this is the case throughout unless explicitly stated otherwise. 

With this assumption, $\maxEW(B)$ can be (somewhat informally) defined as the domain of dependence of the largest quantum-antinormal partial Cauchy slice $c$ with asymptotic boundary $B$ such that any partial Cauchy slice $\tilde c \subset c$ has $\S(\tilde c) \geq \S(c)$ \cite{Akers:2023fqr}. Here quantum-antinormal means that enlarging $c$ slightly cannot increase $\S(c)$ at linear order. The wedge $\minEW(B)$ is defined as the bulk region that is spacelike separated from $\maxEW(\overline{B})$, where $\overline{B}$ is the complement of $B$ in the asymptotic boundary. Thus $\minEW(B)$ is the smallest quantum-normal bulk region such that all information outside it can flow through some partial Cauchy slice to $\overline{B}$.\footnote{With some work, one can show that our simplified definitions of $\maxEW(B)$ and $\minEW(B)$ in terms of von Neumann entropies both reduce to the standard prescription for the entanglement wedge: $\maxEW(B)=\minEW(B)=\EW(B)$. However, in the general case where the max- and min-entropies differ, $\maxEW(B)$ and $\minEW(B)$ may differ as well.} This gives an attractive and intuitive physical picture in which the information flows through the bulk towards the asymptotic boundary, with a maximum information capacity through any given surface controlled by its area.\footnote{Notably, the state is received by an asymptotic \emph{bulk} region $\tilde c$, not by the conformal boundary. This fact aligns well with our proposal that entanglement wedges should be associated to bulk regions, not boundary regions.}

We now return to the problem of associating an entanglement wedge to a gravitating \emph{bulk} region $a$ that need not be static. A careful distinction between a min- and the max-entanglement wedge, $\emin(a)$ and $\emax(a)$, turns out to be critical to this task, even when von Neumann entropies are a good approximation. Indeed, we will see that the two need not agree \emph{even in the $G\hbar\to 0$ limit}, when the generalized entropy $\S(c)$ can be approximated by $\mathrm{Area}(\partial c)/4G\hbar$. Only in static settings (and with the simplifying assumption that von Neumann entropies can be used) will we find that $\emax(a)$ and $\emin(a)$ both reduce to the single prescription given in Ref.~\cite{Bousso:2022hlz}.

\paragraph{Outline} 

In Sec.~\ref{gew} we define the \emph{max- and min-entanglement wedges} of bulk regions, $\emax(a)$ and $\emin(a)$, associated to an arbitrary \emph{wedge}\footnote{We now switch to a more precise formulation, in which the input $a\subset M$ is a wedge, \emph{i.e.}, the maximal causal development of a partial Cauchy slice in $M$.
} $a$ in any gravitating spacetime $M$ that satisfies the semiclassical Einstein equation.

Our definitions of $\emax(a)$ and $\emin(a)$ build on those of $\maxEW(B)$ and $\minEW(B)$, the entanglement wedges of a boundary region $B$. $\emax(a)$ can be thought of as the largest quantum-antinormal region containing $a$, subject to certain modified flow conditions. We will require that information outside of $a$ in $e_{\rm max}(a)$ can flow through a Cauchy slice to the edge of $a$, rather than to a boundary region $B$. And we shall not impose the condition of quantum-antinormality where the edge of $e_{\rm max}(a)$ coincides with the edge of $a$, since no information flows from there. Similarly, $e_{\rm min}(a)$ is the smallest quantum-normal region containing $a$ such that information can flow away from it, across a Cauchy slice of its spacelike complement.

In Sec.~\ref{elementary} we prove that $\emax$ and $\emin$ satisfy the following key properties characteristic of reconstructible regions: 
\begin{itemize}
    \item \emph{Encoding:} Information can flow from $\emax(a)$ toward $a$ through quantum channels whose capacity is controlled by the areas of intermediate homology surfaces. Similarly, quantum information can flow away from $\emin(a)$.
    \item \emph{Inclusion:} $\emin(a)\supset\emax(a)\supset a$.
    \item \emph{No cloning:} $\emax(a)$ is spacelike to $\emin(b)$, if $a$ and $b$ are suitably independent. 
    \item \emph{Nesting:} $\emin(a)\subset\emin(b)$ if $a\subset b$. 
    \item \emph{Strong subadditivity of the generalized entropy:}  $$\S[\emax(a\cup b)]+\S[\emax(b\cup c)]\geq \S[\emax(a\cup b\cup c)]+\S[\emax(b)]~,$$ if $a,b,c$ are mutually spacelike and if $\emax=\emin$ for each of the four sets appearing in the arguments.  
\end{itemize}
These properties mirror those of entanglement wedges of boundary regions in AdS/CFT, $\maxEW(B)$ and $\minEW(B)$. They support the interpretation of $\emax(a)$ as the largest wedge whose semiclassical description can be fully reconstructed from $a$, and of $\emin(a)$ as the complement of the largest wedge about which nothing can be learned from $a$.  

\begin{figure}[]
\begin{subfigure}{.48\textwidth}
  \centering
 \includegraphics[width = 0.75\linewidth]{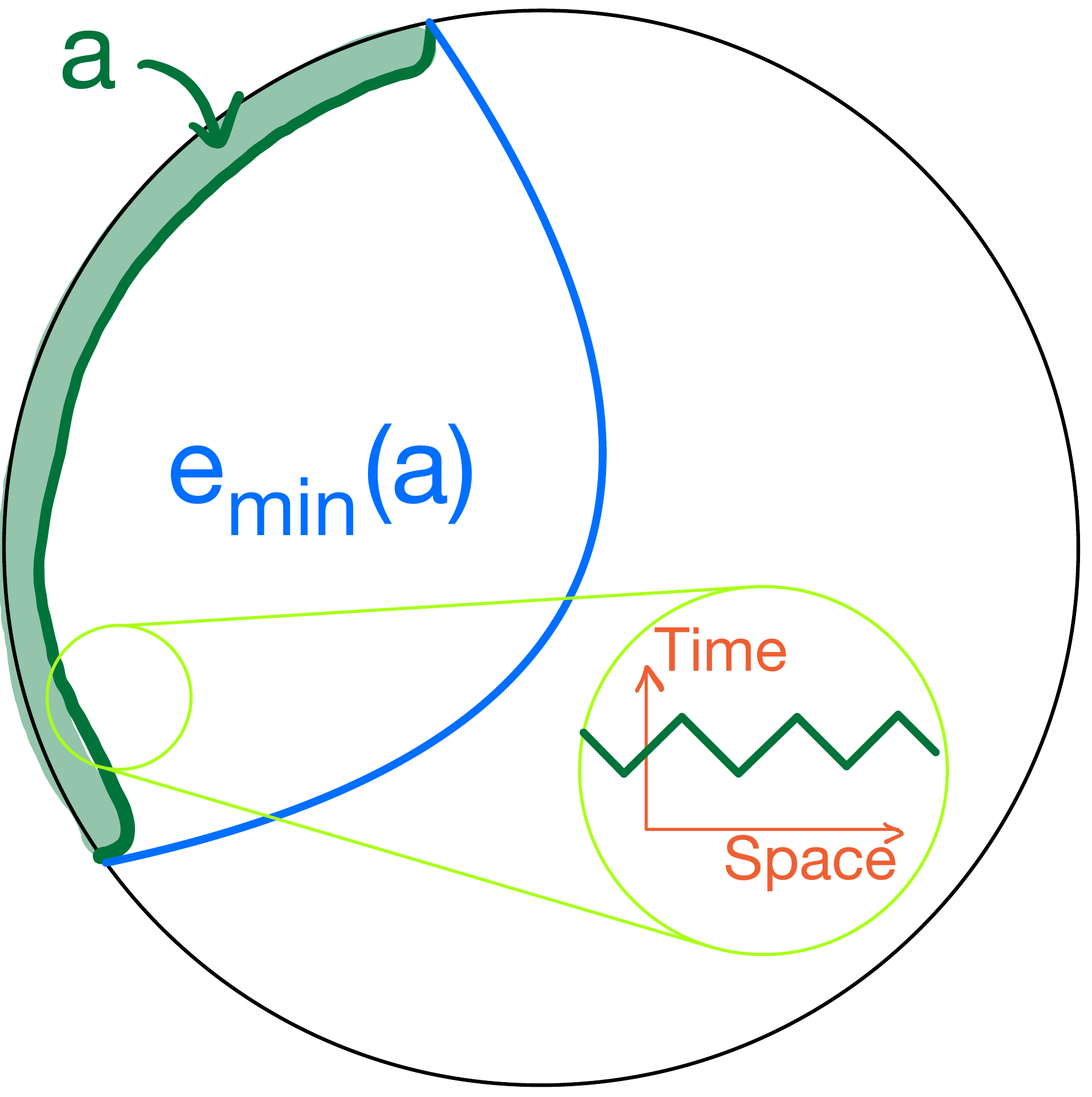}
\end{subfigure}
\begin{subfigure}{.48\textwidth}
  \centering
 \includegraphics[width = 0.75\linewidth]{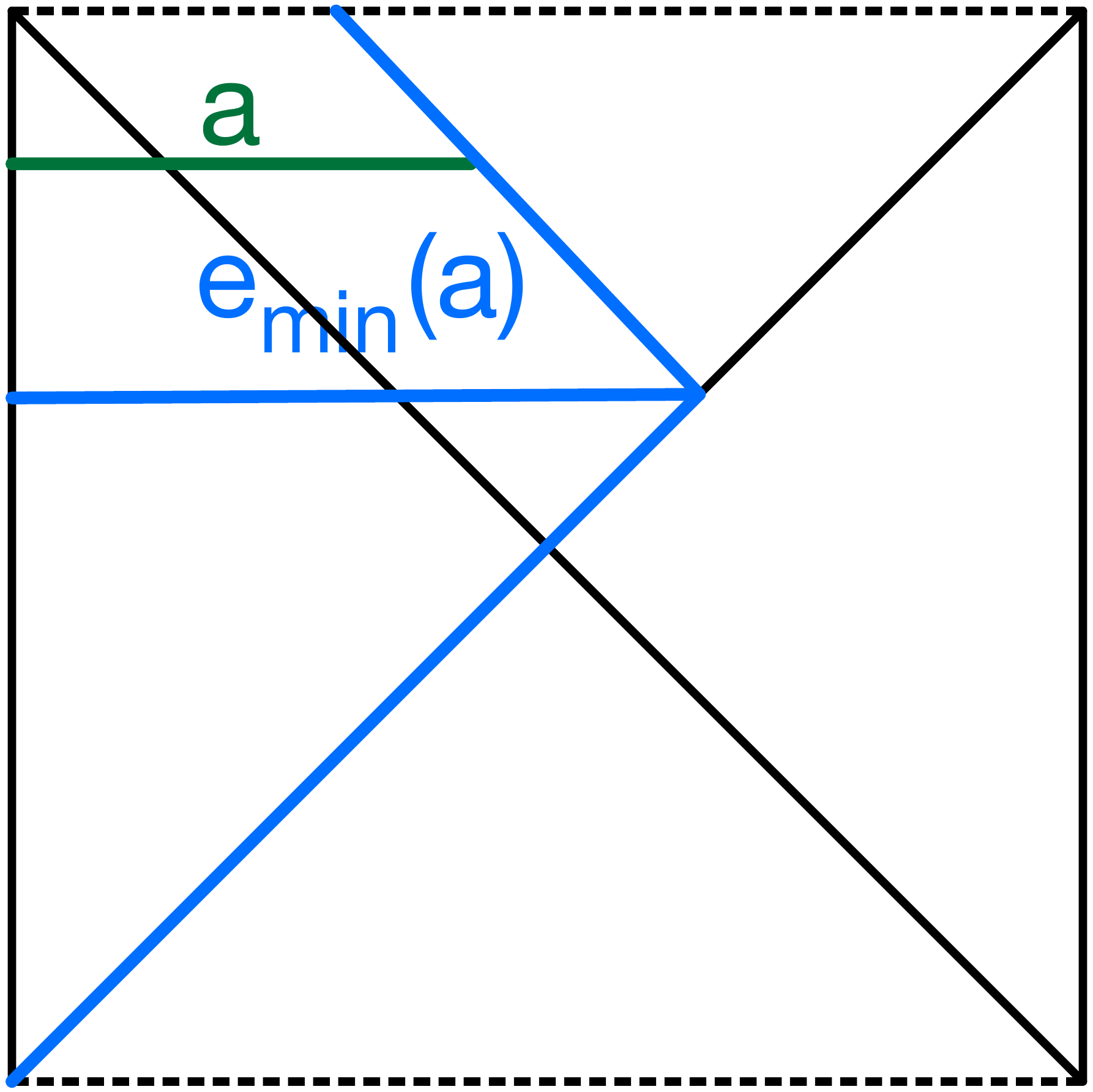}
\end{subfigure}
\caption{The covariant definition of generalized entanglement wedges requires a distinction between max- and min-entanglement wedges. Unlike~\cite{Akers:2020pmf}, these can differ even at the classical level. Two examples are shown; they are chosen asymptotically AdS for ease of drawing, but this is not essential. In both, our definition yields $\emax(a)=a$, while $\emin(a)$ includes $\emax(a)$ as a proper subset. Left: spatial slice of vacuum AdS. The bulk region $a$ has an inner boundary that wiggles up and down in time and so has small area. Right: spacetime diagram of a two-sided black hole. The input region $a$ extends into the black hole interior.}
\label{fig:emin}
\end{figure}

In Sec.~\ref{reduce}, we consider special cases and examples. In Sec.~\ref{static}, we show that $\emax(a)=\emin(a)$ if $a$ lies on a time-reflection symmetric Cauchy slice. In this case our proposal reduces to the much simpler prescription that we had previously formulated for this special case~\cite{Bousso:2022hlz}. In Sec.~\ref{asymptotic}, we show that $\emax(a)$ and $\emin(a)$ reduce to $\maxEW(B)$ and $\minEW(B)$, the max- and min-entanglement wedges of boundary subregions in AdS/CFT, if $a$ is an appropriate asymptotically AdS region with conformal boundary $B$.

In Sec.~\ref{examples}, we construct $\emax$ and $\emin$ explicitly for some examples. Perhaps surprisingly, in some cases $\emax(a)$ will be a proper subset of $\emin(a)$, even though the min- and max-entropies agree with the von Neumann entropies.\footnote{In fact, for brevity and readability, we do not give a fully general definition of $\emax$ and $\emin$ in terms of smooth max- and min-entropies in this paper. This allows us to focus on the challenge of allowing for bulk input regions. It is straightforward to refine our definitions to handle incompressible quantum states, by replacing the von Neumann entropy with max- and min-entropies like in the definitions of $\maxEW$ and $\minEW$~\cite{Akers:2020pmf,Akers:2023fqr}.} They need not coincide even in the classical limit, when the generalized entropy \eqref{eq:sgen} is approximated by the area. Two examples where this happens are shown in Fig.~\ref{fig:emin}.

In Sec.~\ref{discussion}, we discuss some of the implications and challenges that arise from our results.

\section{Max- and Min-Entanglement Wedges of Gravitating Regions}
\label{gew}

Let $M$ be a globally hyperbolic Lorentzian spacetime with metric $g$. The chronological and causal future and past, $I^\pm$ and $J^\pm$, and the future and past domains of dependence and Cauchy horizons, $D^\pm$ and $H^\pm$, are defined as in Wald~\cite{Wald:1984rg}. In particular, the definitions are such that $p\notin I^+(\set{p})$ and $p\in J^+(\set{p})$. Given any set $s\subset M$, $\partial s$ denotes the boundary of $s$ in $M$, and $\cl s\equiv s\cup \partial s$ denotes the closure.

\begin{defn}
The spacelike complement of a set $s\subset M$ is defined by
\begin{equation}
    s'=M\setminus \cl[J^+(s)]\setminus \cl[J^-(s)]~.
\end{equation}
\end{defn}

\begin{figure}[t]
\begin{subfigure}{.48\textwidth}
  \centering
 \includegraphics[width = 0.70\linewidth]{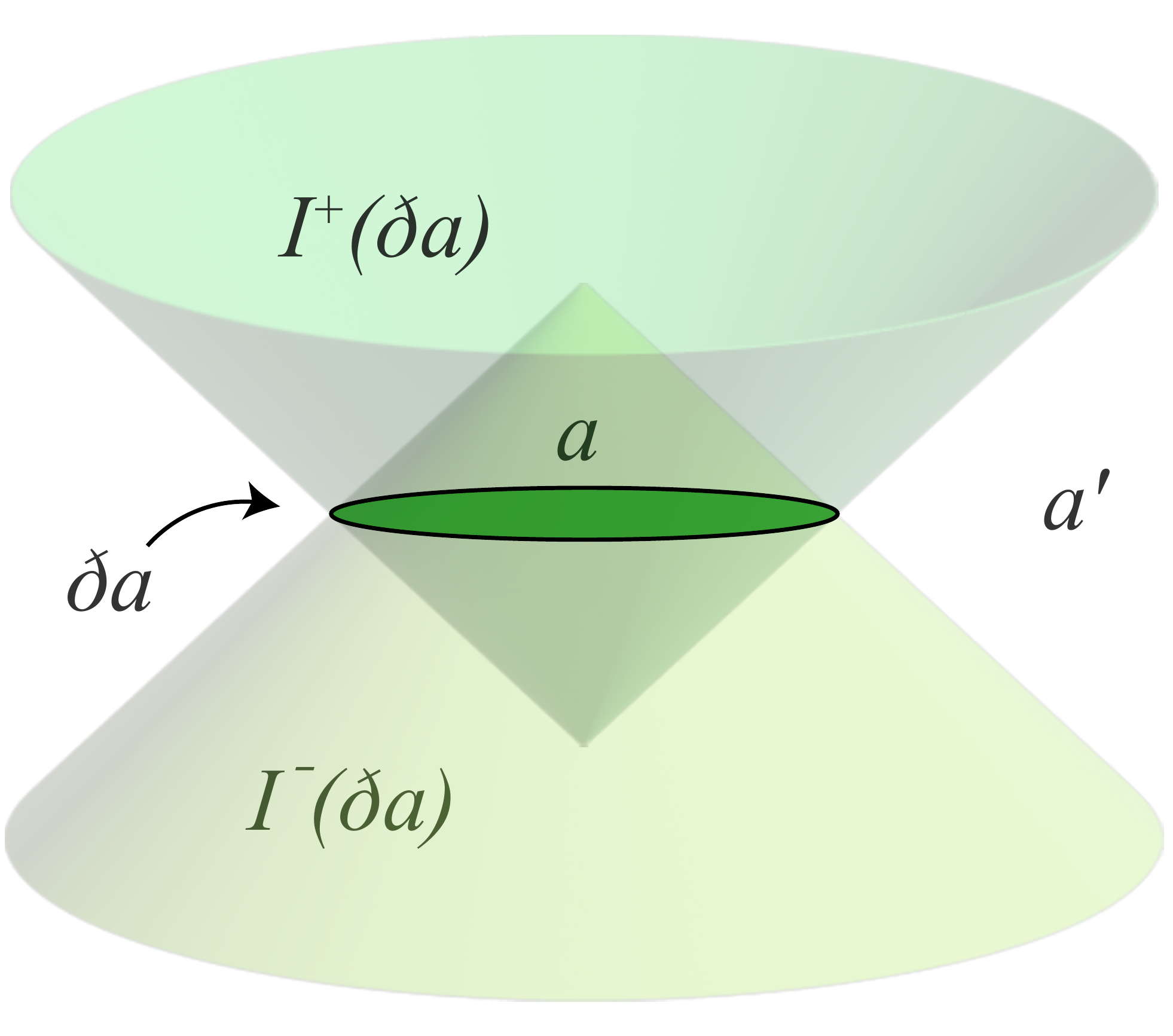}
\end{subfigure}
\begin{subfigure}{.48\textwidth}
  \centering
 \includegraphics[width = 0.9\linewidth]{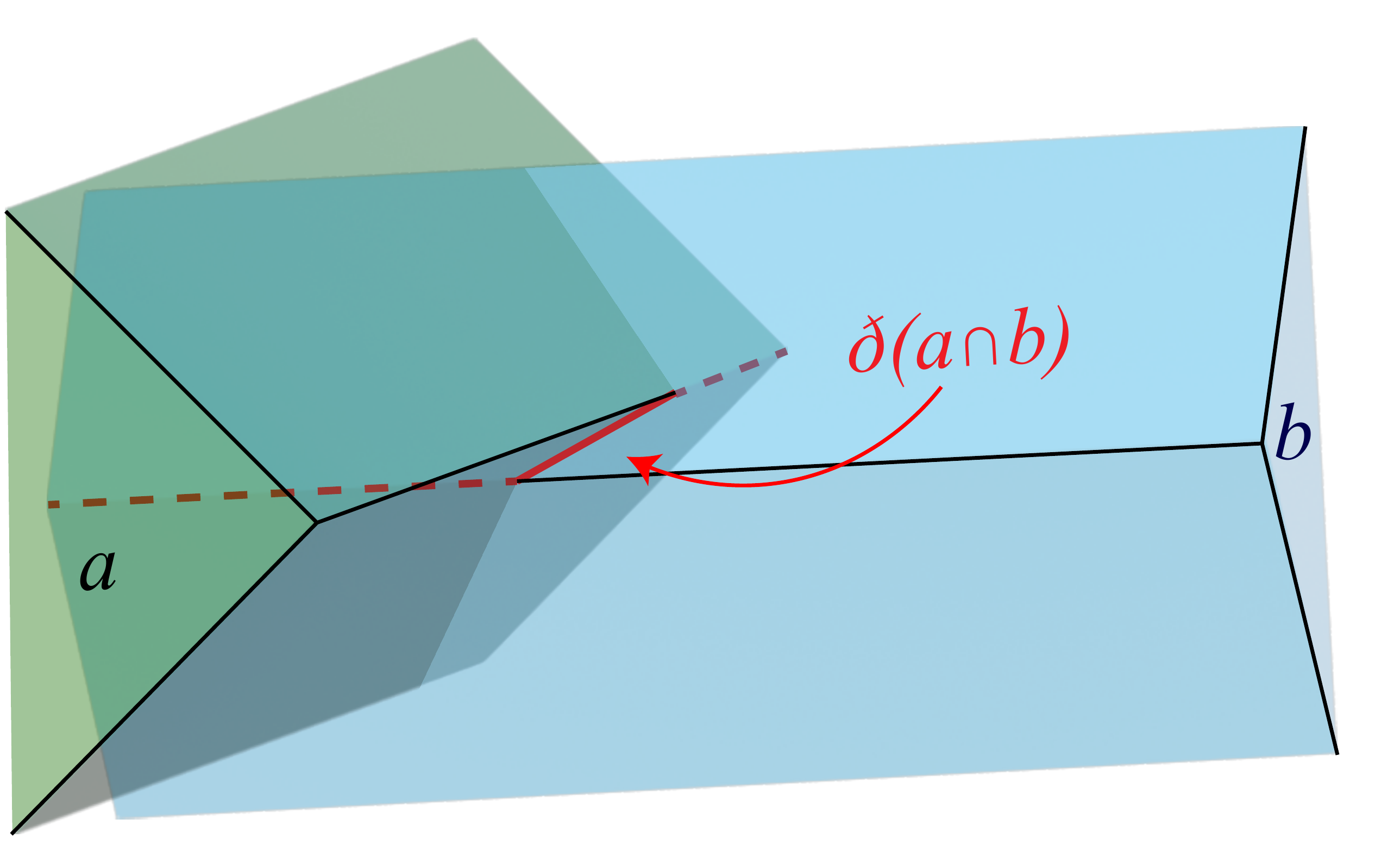}
\end{subfigure}
\caption{\emph{Left:} A wedge $a$ and its complement wedge $a'$ in Minkowski space. Their shared edge $\eth a$ is a sphere. A Cauchy slice of $a$ is shown in dark green. This wedge is ``normal,'' that is, past and future-directed outward lightrays orthogonal to $\eth a$ expand.  \emph{Right:} The intersection of two wedges is again a wedge. Its edge decomposes as $\eth(a \cap b) = \cl\left[(\eth a\cap b) \sqcup (H^+(a)\cap H^-(b))\sqcup \set{a\leftrightarrow b}\right]$.}
\label{fig-wedges}
\end{figure}

\begin{defn}\label{def:covwedge}
A {\em wedge} is a set $a\subset M$ that satisfies $a=a''$ (see Fig.~\ref{fig-wedges}, left). 
\end{defn}

\begin{rem}\label{wilem}
The intersection of two wedges $a,b$ is easily shown to be a wedge (see Fig. \ref{fig-wedges}, right); similarly, the \emph{complement wedge} $a'$ is a wedge:
\begin{equation}
    (a\cap b)'' = a\cap b~;~~a'''=a'~.
\end{equation}
\end{rem}

\begin{figure}[t]
\begin{center}
  \includegraphics[width=0.8\linewidth]{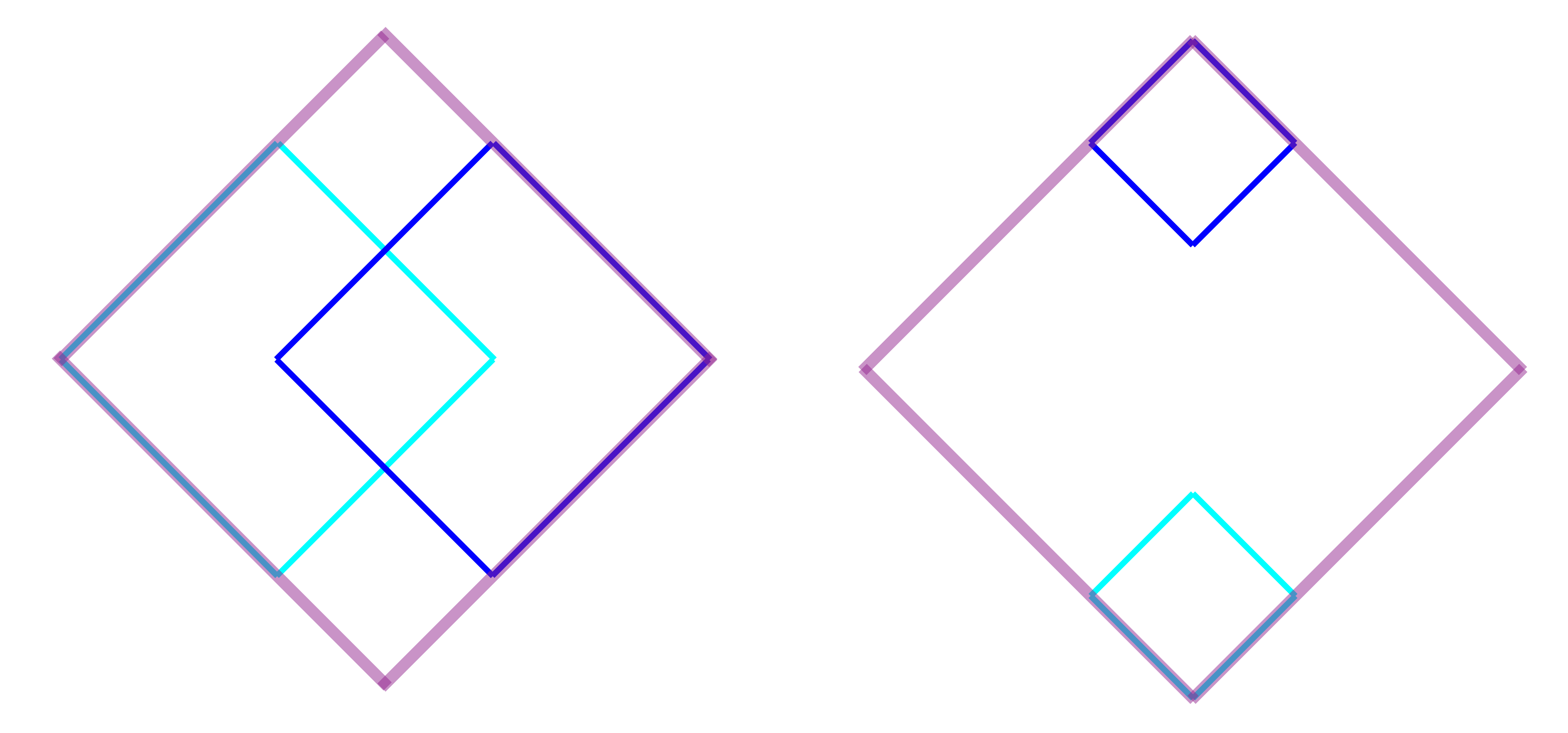} 
\end{center}
\caption{The wedge union (purple) of two wedges (turquoise, blue) is the smallest wedge that contains both. Two examples are shown.}
\label{fig:union}
\end{figure}

\begin{defn}
Given two wedges $a$ and $b$, we define the {\em wedge union} as $a\Cup b\equiv (a'\cap b')'$ (see Fig. \ref{fig:union}). By the above remark, $a\Cup b$ is always a wedge. 
\end{defn}
\begin{defn}\label{def:wedgeuniongen}
    For notational simplicity, it will be convenient to extend this definition to sets $s,t$ that need not be wedges: $s\Cup t\equiv s''\Cup t''$.
\end{defn}

\begin{rem}
The wedge union satisfies $a\Cup b \supset a\cup b$. It is the smallest such wedge: any wedge that contains $a\cup b$ will contain $a\Cup b$.
\end{rem}
\begin{defn}
The {\em edge} $\eth a$ of a wedge $a$ is 
defined by $\eth a\equiv \partial a \cap \partial a'$. Conversely, a wedge $a$ can be fully characterized by specifying its edge $\eth a$ and one spatial side of $\eth a$.
\end{defn}

\begin{defn}
The \emph{generalized entropy}~\cite{Bekenstein:1972tm} of a wedge $a$ is defined as
\begin{equation}\label{sgendef}
    \S(a) \equiv \frac{\A(\eth a)}{4G\hbar} + S(a)+ \ldots~,
\end{equation}
where $S$ is the von Neumann entropy of the reduced quantum state of the matter fields on any Cauchy slice of $a$. The ellipsis stands for additional gravitational counter-terms~\cite{Bousso:2015mna} that cancel subleading divergences in $S(a)$. For a set $s \subset M$ that is not a wedge, we will sometimes write $\S(s) = \S(s'')$ to simplify notation.
\end{defn}

\begin{conv} \label{conv:globalpurity}
We assume throughout that the global state is pure, and hence that
\begin{align}
    \S(a) = \S(a')
\end{align}
for all wedges $a$. If necessary, this can be achieved by purifying the state using an external system $R$, and including $R \subset a'$ whenever $R \not\subset a$.
\end{conv}

\begin{defn}\label{def:conformaledge}
Given a wedge $a$, we distinguish between its edge $\eth a$ in $M$ and its edge $\deltabar a$ as a subset of the conformal completion (also called Penrose diagram or unphysical spacetime) $\tilde M$~\cite{Wald:1984rg}. If $a$ is asymptotic, the latter set may contain an additional piece, the {\em conformal edge}
\begin{equation}
    \tilde\eth a\equiv \deltabar a\cap \partial\tilde M~.
\end{equation}
\end{defn}

\begin{defn}
Let $a$ be a wedge, and let $p\in \eth a$. The future (past) {\em expansion}, $\Theta^+(a,p)$ ($\Theta^-(a,p)$), is the shape derivative of the generalized entropy under outward deformations of $a$ along the future (past) null vector field orthogonal to $\eth a$ at $p$:\footnote{More commonly, $\Theta^\pm$ is referred to as a \emph{quantum expansion}, in order to distinguish it from the classical expansion $\theta$. We omit the modifier for brevity.\label{noquantum}}
\begin{equation}
    \Theta^\pm(a,p)\equiv 4G\hbar \, \frac{\delta S_{\rm gen}[a(X^\pm(p))]}{\delta X^\pm(p)}~.
\end{equation}
Here $X^+$ ($X^-$) is an affine parameter for the future (past) null congruence orthogonal to $\eth a$; and $a(X^\pm(p))$ are wedges obtained by deforming $\eth a$ along these congruences. See Refs.~\cite{Bousso:2015mna,Leichenauer:2017bmc} for further details.
\end{defn}

\begin{rem}
By Eq.~\eqref{sgendef}, the expansion can be decomposed as
\begin{equation} \label{eq:qexpansion}
    \Theta^\pm(a,p) = \theta^\pm(p)+4G\hbar\, \frac{\delta S[a(X^\pm(p))]}{\delta X^\pm(p)}~.
\end{equation}
The first term is the classical expansion~\cite{Wald:1984rg}, which depends only on the shape of $\eth a$ near $p$. The second term in Eq.~\eqref{eq:qexpansion} is nonlocal.
\end{rem}



\begin{defn}
The wedge $a$ is called \emph{normal}\,\footnote{Again we drop the common modifier ``quantum-'' (\emph{quantum-normal} etc.)\ because it would become too cumbersome; see footnote~\ref{noquantum}.} at $p\in \eth a$ if $\Theta^+(a,p)\geq 0$ and $\Theta^-(a,p)\geq 0$. For other sign combinations, $a$ is called \emph{antinormal} ($\leq,\leq$), \emph{trapped} ($\leq,\geq$), \emph{anti-trapped} ($\geq,\leq$) and \emph{ extremal} ($=,=$) at $p$. Marginal cases arise if one expansion vanishes at $p$. In relations that hold for all $p\in \eth a$, we omit the argument $p$: if $\Theta^+(a)\geq 0$ and $\Theta^-(a)\geq 0$, we simply call $a$ \emph{normal}; and similarly for the other cases.
\end{defn}

\begin{conj}[Quantum Focussing Conjecture] \label{conj-qfc}
The quantum expansion is nonincreasing along a null congruence~\cite{Bousso:2015mna}:
\begin{equation}\label{qfc}
   \frac{\delta \Theta^\pm(a,p)}{\delta X^\pm(\bar{p})}\leq 0~.
\end{equation}
\end{conj}
\begin{rem}
When the functional derivatives are taken at different points, $p\neq \bar{p}$, the Quantum Focussing Conjecture (QFC) follows from strong subadditivity~\cite{Bousso:2015mna}. The QFC implies the Quantum Null Energy Condition~\cite{Bousso:2015mna} in a certain nongravitational limit; the latter statement can be proven within quantum field theory for both free and interacting theories~\cite{Bousso:2015wca, Balakrishnan:2017bjg}; see also Refs.~\cite{Koeller:2015qmn, Wall:2017blw, Ceyhan:2018zfg, Balakrishnan:2019gxl}. The ``diagonal'' case of the QFC, $p=\bar{p}$, remains a conjecture. A weaker version~\cite{Shahbazi-Moghaddam:2022hbw} suffices for all proofs in the present paper and has been proven holographically for brane-worlds in AdS.
\end{rem}

\begin{conv}[Genericity Condition]
The inequality Eq.~\eqref{qfc} is generically strict. We will assume this stronger condition whenever necessary.
\end{conv}

\begin{defn}[Max-Entanglement Wedge of a Gravitating Region]\label{emaxdef}
  Given a wedge $a$, let $F(a)\equiv \set{f:\mathrm{I}\,\wedge\,\mathrm{II}\,\wedge\,\mathrm{III}}$ be the set of all wedges that satisfy the following properties:
  \begin{enumerate}[I.]
  \item $f\supset a$ and $\tilde \eth f=\tilde \eth a$;
  \item $f$ is antinormal at points $p\in \eth f\setminus\eth a$;
  \item $f$ admits a Cauchy slice $\Sigma$ such that
    \begin{enumerate}
    \item $\Sigma\supset \eth a$;
    \item $\S(h) > \S(f)$ for any wedge $h \neq f$ such that $a\subset h$, $\eth h\subset \Sigma$, and $\eth h\setminus \eth f$ is compact in $M$.
    \end{enumerate}
   \end{enumerate}
  The {\em max-entanglement wedge} of $a$, $\emax(a)$, is their wedge union:
   \begin{equation}\label{eq:emaxdef}
       \emax(a) \equiv \Cup_{f\in F(a)}\, f~.
   \end{equation}
\end{defn}

\begin{defn}[Min-Entanglement Wedge of a Gravitating Region]\label{emindef}
  Given a wedge $a$, let $G(a)\equiv \set{g: \mathrm{i}\, \wedge\, \mathrm{ii} \,\wedge\, \mathrm{iii}}$ be the set of all wedges that satisfy the following properties:
  \begin{enumerate}[i.]
  \item $g\supset a$ and $\tilde \eth g=\tilde \eth a$;
  \item $g$ is normal; 
  \item $g'$ admits a Cauchy slice $\Sigma'$ such that $\S(h) > \S(g)$ for any wedge $h \neq g$ such that $g \subset h$, $\eth h\subset \Sigma'$, and $\eth h\setminus \eth g$ is compact.
  \end{enumerate}
  The {\em min-entanglement wedge} of $a$, $\emin(a)$, is their intersection:
  \begin{equation}\label{eq:emindef}
      \emin(a)\equiv \cap_{g\in G(a)}\, g~.
  \end{equation}
\end{defn}

\section{Properties}
\label{elementary}

\begin{thm}\label{lem:emaxpropshare}
$\emax(a)\in F(a)$.
\end{thm}

\begin{proof} 
We must show that $\emax(a)$ satisfies properties I--III listed in Def.~\ref{emaxdef}.

\emph{Property I:} $f=a$ satisfies properties I--III with any choice of Cauchy slice. Hence $F(a)$ is nonempty, and Eq.~\eqref{eq:emaxdef} implies that $f=\emax(a)$ satisfies property I.

\emph{Property II:} For any $p \in \eth f_3 \setminus \eth a$, either (a) $p \in \eth f_1 \cup \eth f_2$ or (b) $p \in H^+(f_1') \cap H^-(f_2')\cup H^-(f_1') \cap H^+(f_2')$. Since $f_1$ and $f_2$ themselves satisfy property II, property II for $f_3$ follows by either (a) strong subadditivity, or (b) the Quantum Focusing Conjecture~\ref{conj-qfc} followed by strong subadditivity. By induction, $\emax(a)$ is antinormal at points $p \in \eth \emax(a) \setminus \eth a$.

\emph{Property III:} Again proceeding inductively, let $f_1,f_2\in F(a)$, with property III satisfied by Cauchy slices $\Sigma_1$ and $\Sigma_2$, respectively. Then $f_3=f_1\Cup f_2$ admits the Cauchy slice
\begin{equation}\label{eq:unionslice}
  \Sigma_3 = \Sigma_1 \cup  [H^+(f_1')\cap \mathbf{J}^-(\Sigma_2)] 
  \cup [H^-(f_1')\cap \mathbf{J}^+(\Sigma_2)] \cup [\Sigma_2 \cap f_1']~.
\end{equation}
$\Sigma_3$ trivially satisfies property IIIa. To prove IIIb, let $h\supset a$, $\eth h\subset \Sigma_3$. By property IIIb of $\Sigma_1$, the QFC, and property IIIb of $\Sigma_2$, respectively,\footnote{Here we are using the notation from Def.~\ref{def:wedgeuniongen} for the wedge union $s \Cup t = (s'') \Cup (t'')$ of arbitrary sets $s,t$ that are not necessarily wedges.}
\begin{equation}\label{eq:uniongen}
  \S(h)\geq \S(h\Cup \Sigma_1)\geq \S[h \Cup (\Sigma_3\setminus \Sigma_2)]\geq
  \S[\Sigma_3]~.
\end{equation}
Strong subadditivity was used in every step.\footnote{The QFC does not imply that $\S$ decreases along portions of $H(f_1')$ that originate at $a$. Accordingly, these are added only in the last step. Separately, we note that our construction would establish $f_3\in F(a)$ even if $f_2$ did not satisfy the requirement that $\Sigma_2\supset \eth a$; this fact will be important in the proof of Theorem~\ref{thm:emaxtomaxEW}.\label{asymmetric}} At least one of these inequalities is strict whenever $h \neq f_3$.
\end{proof}

\begin{conv}
Let $$\smax(a)$$ denote a Cauchy slice of $\emax(a)$ that satisfies property III, and which exists by the preceding Lemma. 
\end{conv}

\begin{thm}\label{thm:emaxextremal}
All antinormal portions of $\eth\emax(a)$ are extremal. In particular, $\emax(a)$ is extremal at points $p\in\eth\emax(a)\setminus\eth a$.
\end{thm}
\begin{proof}
Suppose that $\emax(a)$ is antinormal at $p$; that is $\Theta^\pm[\emax(a),p]\leq 0$. Suppose for contradiction (and without loss of generality) that $\Theta^-[\emax(a),p]<0$. Continuity of $\Theta^-$ and the QFC for $\Theta^+$ imply that $\eth \emax(a)$ can be deformed along the outward future-directed null congruence at $p$ to generate a wedge $f_1$ that still satisfies properties I--III of Def.~\ref{emaxdef}. Since $f_1 \not\subset \emax(a)$, this contradicts the definition of $\emax(a)$.

The second part of the theorem follows since $\emax(a)$ is antinormal at $p\in\eth\emax(a)\setminus\eth a$, by Theorem~\ref{lem:emaxpropshare}.
\end{proof}

\begin{cor}
    No point on $\eth \emax(a)$ can be properly null separated from $\eth a$; that is:
    \begin{equation}
        \eth\emax(a)\cap H(a') \subset \eth a~.
    \end{equation}
\end{cor}

\begin{proof}
    If such a point $p$ existed, any Cauchy slice of $\emax(a)$ that contains $\eth a$ would contain a null geodesic orthogonal to $\eth\emax(a)$ at $p$. Local inward deformations at $p$ along this geodesic decrease the generalized entropy, by the preceding theorem and the QFC, in contradiction with property III of $\emax$ established in Theorem~\ref{lem:emaxpropshare}.
\end{proof}

\begin{thm}\label{lem:eminpropshare}
$\emin(a)\in G(a)$. 
\end{thm}

\begin{proof} 
We must show that $\emin(a)$ satisfies properties i--iii listed in Def.~\ref{emindef}.

\emph{Property i:} $g=M$ trivially satisfies properties i--iii, so $G(a)$ is nonempty. Property i then implies $\emin(a)\supset a$.

\emph{Property ii:} The intersection of two normal wedges is normal by Lemma 4.14 of Ref.~\cite{Bousso:2022hlz}. Hence $\emin(a)$ is normal by Def.~\ref{emindef}.

\emph{Property iii:} Let $g_1,g_2\in G(a)$ with property iii satisfied by Cauchy slices $\Sigma'_1$ and $\Sigma'_2$, respectively; and let $g_3=g_1\cap g_2$. Then $g_3'$ admits the Cauchy slice
\begin{equation}\label{eq:sminunion}
  \Sigma'_3 = \Sigma'_1 \cup  [H^+(g_1)\cap \mathbf{J}^-(\Sigma'_2)] 
  \cup [H^-(g_1)\cap \mathbf{J}^+(\Sigma'_2)] \cup [\Sigma'_2 \cap g_1]~.
\end{equation}
Let $h\supset g_3$, $\eth h\subset \Sigma'_3$. By property iii of $\Sigma'_1$, the QFC, and property iii of $\Sigma'_2$, respectively,
\begin{equation}\label{eq:sminunionworks}
  \S(h')\geq \S(h'\Cup \Sigma_1')\geq \S[h' \Cup (\Sigma_3'\setminus \Sigma_2')]\geq
  \S[\Sigma'_3]~.
\end{equation}
Strong subadditivity was used for each inequality. At least one of these inequalities is strict whenever $h \neq g_3$. Hence, using Convention \ref{conv:globalpurity}, we have $\S(h) > \S(g_3)$. Property iii follows by induction.
\end{proof}

\begin{conv}
Let $$\smin(a)$$ denote a Cauchy slice of $\emin'(a)$ that satisfies property iii, and which exists by the preceding Lemma. 
\end{conv}

\begin{thm}\label{thm:eminmarginal}
$\emin(a)$ is marginal or extremal at points $p\in \eth \emin(a)\setminus\eth a$. Specifically,
\begin{itemize}
    \item $\emin(a)$ is marginally anti-trapped, $\Theta^-[\emin(a),p]=0$, at $p\in \eth \emin(a) \cap H^+(a')$.
    \item $\emin(a)$ is marginally trapped, $\Theta^+[\emin(a),p]=0$, at $p\in \eth \emin(a) \cap   H^-(a')$.
    \item $\emin(a)$ is extremal, $\Theta^+[\emin(a),p]=\Theta^-[\emin(a),p]=0$, at $p\in \eth \emin(a) \cap a'$.
\end{itemize}
\end{thm}

\begin{proof}
Suppose that $p\in H^+(a')$, and let $\xi$ be the open geodesic segment connecting $\eth a$ to $p$. For $q \in \xi$ let $\tilde e_{\rm min}(a)$ be defined by locally deforming $\eth \emin(a)$ along $H^-[\emin(a)]$ near $p$ such that $q \in \tilde e_{\rm min}(a)$.  By Theorem~\ref{lem:eminpropshare}, $\Theta^+[\emin(a),p]\geq 0$ and hence, by the QFC, $\Theta^+[\tilde e_{\rm min}(a),q]\geq 0$ and $\S[\tilde e_{\rm min}(a)] \leq \S(\emin(a))$. This conflicts with the definition of $\emin(a)$ unless $\Theta^-$ immediately becomes negative under this deformation. By continuity, $\Theta^-[\emin(a),p]=0$. The time-reversed argument applies to $p\in H^-(a')$.

For $p \in a'$, we similarly deform $\eth \emin(a)$ inward (\emph{i.e.}, towards $a$) at $p$ along a past null geodesic to show that $\Theta^-[\emin(a),p]=0$, and inward along a future null geodesic to show that $\Theta^+[\emin(a),p]=0$. 
\end{proof}


\begin{thm}\label{thm:emaxemin}
$\emax(a)\subset\emin(a)$.
\end{thm}

\begin{figure}[t]
\begin{subfigure}{.48\textwidth}
  \centering
 \includegraphics[width = 0.8\linewidth]{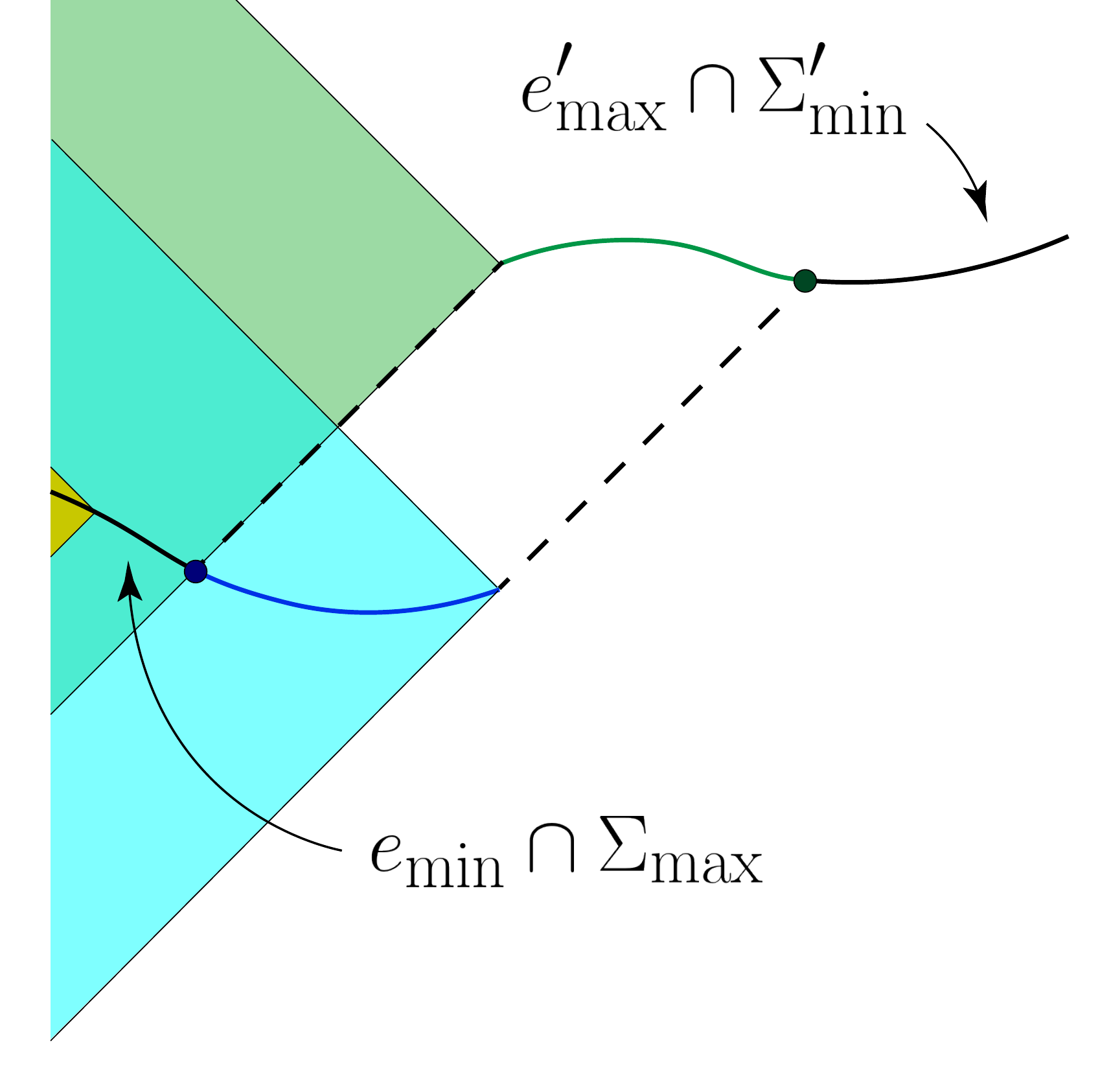}
\end{subfigure}
\begin{subfigure}{.48\textwidth}
  \centering
 \includegraphics[width = 0.85\linewidth]{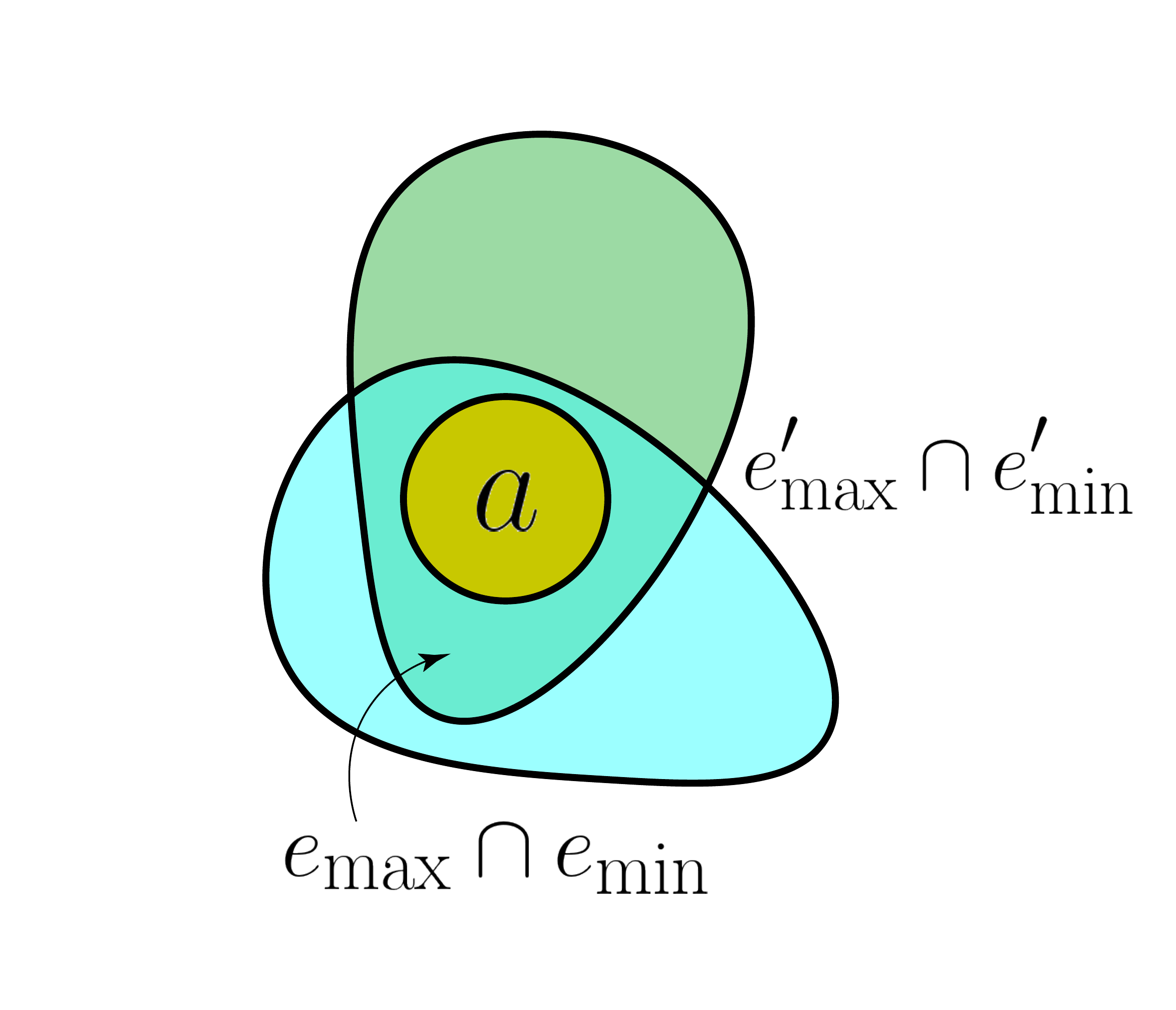}
\end{subfigure}
\caption{Two special cases of the proof of Theorem~\ref{thm:emaxemin}, that $\emax \subset \emin$. In both, we assume that $\emax$ (blue) is not contained in $\emin$ (green) and derive a contradiction. \emph{Left:} In two spacetime dimensions, we have $\S[\emin] \geq \S[(\emin \cap \smax)$ and $\S[\emax] \geq \S[(\emax' \cap \smin)$ by quantum focussing. But $\S[(\emin \cap \smax) \geq \S[\emax]$ and $\S[(\emax' \cap \smin) \geq \S[\emin]$ by properties IIIb and iii respectively. The inequalities are strict, and hence we have a contradiction, unless $\emax \subset \emin$. \emph{Right:} If there exists a Cauchy slice $\Sigma$ (shown in figure) such that $\smax,\smin \in \Sigma$, then properties IIIb and iii imply $\S[\emax \cap \emin] \geq \S[\emax]$ and $\S[\emax' \cap \emin'] \geq \S[\emin]$ with strict inequalities unless $\emax \subset \emin$. But $\S[\emax] + \S[\emin] \geq \S[\emax \cap \emin] + \S[\emax' \cap \emin']$ by strong subadditivity. The general proof involves combining the techniques used for these two special cases.}
\label{fig:emaxinemin}
\end{figure}

\begin{proof}
Two special cases of this proof are illustrated in Fig.~\ref{fig:emaxinemin}. Since only one input wedge $a$ is involved, we suppress the argument of $\emax$ and $\emin$. normality of $\emin$ and $\emax'$ (outside of $\emin)$ implies
\begin{align}\label{eq:emaxinemin}
    \S(\emin)+\S(\emax) & \nonumber
    \geq \S[(\emin \cap \smax) \Cup (\emin \cap \emax')] \\ & \hspace{5em}
         + \S[(\emax' \cap \smin) \Cup (\emin \cap \emax')] \\ &
    \geq \S(\emin \cap \smax) + \S(\emax' \cap \smin)~.
\end{align}
To obtain the second inequality, note that the area of the relevant edge portions decreases or remains constant, and the von Neumann entropy obeys strong subadditivity.

But $\emin \cap \smax$ satisfies the conditions demanded of $h$ in property IIIb, Def.~\ref{emaxdef}. And $\emax' \cap \smin$ satisfies the conditions demanded of $h$ in property iii, Def.~\ref{emindef}. The definitions imply
\begin{equation}
  \S(\emin \cap \smax) + \S(\emax' \cap \smin) \geq \S(\emin)+\S(\emax)~.
\end{equation}
This inequality is strict and we have a contradiction unless $\emin \cap \smax=\smax$ (and hence, $\emax' \cap \smin=\smin$).
\end{proof}



\begin{thm}[Nesting of $\emin$]\label{thm:nesting}
For wedges $a$ and $b$,
\begin{equation}
    a\subset b \implies \emin(a)\subset \emin(b)~.
\end{equation}
Moreover, $\smin(a)$ can be chosen so that
\begin{equation}
    \smin(a)\supset\smin(b)~.
\end{equation}
\end{thm}

\begin{proof}
By Theorem~\ref{lem:eminpropshare}, $\emin(b)$ satisfies properties i--iii of $f_\alpha(a)$ enumerated in Def.~\ref{emindef}. By Eq.~\eqref{eq:emindef}, $\emin(a)\subset \emin(b)$.

Suppose now that $\smin(a)\not\supset \smin(b)$. Then set $\tilde\Sigma = \smin(a)$ and redefine
\begin{equation}
    \smin(a)\equiv \smin(b) \cup \left( H^+[\emin(b)]\cap \mathbf{J}^-(\tilde\Sigma)\right)
    \cup \left( H^-[\emin(b)]\cap \mathbf{J}^+(\tilde\Sigma)\right)
    \cup [\tilde\Sigma\cap \emin(b)]~.
\end{equation}
This satisfies property iii for $\emin(a)$, by arguments that parallel those given in support of Eq.~\eqref{eq:sminunionworks}.
\end{proof}

\begin{cor} \label{cor:monoemin}
If $a\subset b$ and $\eth \emin(b)\setminus \eth \emin(a)$ is compact, then
\begin{equation}
    \S[\emin(a)]\leq \S[\emin(b)]~.
\end{equation}
\end{cor}

\begin{proof}
By the previous theorem, we can take $\smin(a)\supset\smin(b)$. By compactness of $\eth \emin(b)\setminus \eth \emin(a)$, we may now invoke property iii for $\smin(a)$, with the choice $h=\emin(b)$.
\end{proof}

\begin{thm}[No Cloning]\label{thm:nocloning}
   \begin{equation}
     a \subset \emin'(b)~\mbox{and}~b \subset \emax'(a)~\implies~
     \emax(a)\subset\emin'(b)~.
  \end{equation}
\end{thm}

\begin{proof}
Let 
\begin{equation}
    g=\emin(b)\cap \emax'(a)~.
\end{equation}
We will show that $g$ satisfies properties i-iii listed in Def.~\ref{emindef}. This contradicts the definition of $\emin(b)$ unless $g=\emin(b)$, which is equivalent to the conclusion.

Property i: By assumption, $b\subset \emax'(a)$. By Theorem~\ref{lem:eminpropshare}, $b\subset \emin(b)$. Hence $b\subset g$.

Property ii: By Theorem~\ref{lem:eminpropshare}, $\emin(b)$ is normal. By assumption, $\emin(b)\cap a=\varnothing$, so by Theorem~\ref{thm:emaxextremal}, the null geodesics connecting $\eth\emax(a)$ to $\eth g\cap H[\emax'(a)]$ originate from points where $\emax(a)$ is extremal and thus normal. By the arguments in the proof of Lemma 4.14 of Ref.~\cite{Bousso:2022hlz}, $g$ is normal. 

Property iii: Let 
\begin{multline}
    \Sigma' = \smin(b) 
    \cup \left( H^+[\emin(b)] \cap \mathbf{J}^-[\smax(a)]\right) \\
    \cup \left( H^-[\emin(b)] \cap \mathbf{J}^+[\smax(a)]\right)
    \cup [\smax(a) \cap \emin(b)]~.
\end{multline}
This is a Cauchy slice of $g'$. Let $h\supset g$ with $\eth h\subset \Sigma'$. By property iii of $\smin(b)$, the QFC, and property III of $\smax(a)$, respectively,
\begin{equation}
  \S(h')\geq \S[h'\Cup \smin(b)]\geq \S[ h' \Cup (\Sigma'\setminus \smax(a))]
  \geq \S[\Sigma']~.
\end{equation}
Strong subadditivity was used for each inequality. At least one of these inequalities is strict whenever $h \neq g$. Hence, by Convention \ref{conv:globalpurity}, we have $\S(h) > \S(g)$.
\end{proof}

\begin{rem}
The following theorem suggests that when $\emin(a)=\emax(a)$, the generalized entropy of the entanglement wedge is the von Neumann entropy of the quantum state corresponding to the fundamental description of the entanglement wedge.
\end{rem}

\begin{thm}[Strong Subadditivity of the Generalized Entropy]\label{thm:ssa}
Suppressing $\Cup$ symbols where they are obvious, let $a$, $b$, and $c$ be mutually spacelike wedges, such that
\begin{align}\nonumber
    \emin(ab) &=\emax(ab)~,~\emin(bc)=\emax(bc)~,\\ \nonumber
    \emin(b) &=\emax(b)~,~\mbox{and}~\emin(abc)=\emax(abc)~.
\end{align}
Then (writing $e$ for $\emin=\emax$)
\begin{equation}
    \S[e(ab)]+\S[e(bc)]\geq \S[e(abc)]+\S[e(b)]~.
\end{equation}
\end{thm}

\begin{proof}
We define the wedge $x$ by the Cauchy slice of its complement:
\begin{equation}
    \Sigma'(x) = \Sigma'(ab) \cup
    \left( H^+[e(ab)] \cap \mathbf{J}^-[\eth e(bc)]\right) \cup
    \left( H^-[e(ab)] \cap \mathbf{J}^+[\eth e(bc)]\right)~.
\end{equation}
normality of $e(ab)$ and the QFC imply
\begin{equation}
    \S[e(ab)]\geq \S(x)~.
\end{equation}

Note that $\eth x$ is nowhere to the past or future of $\eth e(bc)$. Therefore, there exists a single Cauchy slice that contains the edges of $x$, $e(bc)$, $x\cap e(bc)$, and $x\Cup e(bc)$. Strong subadditivity of the von Neumann entropy implies
\begin{equation}
    S(x)+S[e(bc)]\geq S[x\cap e(bc)]+S[x\Cup e(bc)]~.
\end{equation}
The areas of edges obey the analogous inequality, so
\begin{equation}
    \S(x)+\S[e(bc)]\geq \S[x\cap e(bc)] + \S[x\Cup e(bc)]~.
\end{equation}

Note also that $x\cap e(bc)=e(ab)\cap e(bc)$ and $x\Cup e(bc) \subset e(ab)\Cup e(bc)$. By Theorem~\ref{thm:nesting},
\begin{equation}
    x\Cup e(bc) \subset e(abc)~~\mbox{and}~~x\cap e(bc) \supset e(b)~.
\end{equation}

By Lemma 4.14 of Ref.~\cite{Bousso:2022hlz}, $x\cap e(bc)$ is normal, and $x\Cup e(bc)$ is antinormal except at points where its edge coincides with $\eth(abc)$ and hence with $\Sigma(abc)$. Hence 
\begin{align}
    \S(x\Cup e(bc)) & \geq \S[\Sigma(abc)\setminus (x' \cap e(bc)')]\geq \S[e(abc)]\\
    \S(x\cap e(bc)) & \geq \S[e(b) \Cup (\Sigma'(b)\cap x\cap e(bc))]\geq \S[e(b)]~.
\end{align}
The last inequality in each line follows from properties III and iii of $e(abc)$ and $e(b)$, respectively, as established by  Theorem~\ref{lem:emaxpropshare} and Theorem~\ref{lem:eminpropshare}.
\end{proof}

\section{Special Cases and Examples}
\label{reduce}

\subsection{Time-reversal Invariant Case}
\label{static}

\begin{defn}\label{ewstatic}
Let $M$ be a time-reflection symmetric spacetime. That is, $M$ admits a $\mathbb{Z}_2$ symmetry generated by an operator $T$ that exchanges past and future. Let $\Sigma_0$ be the Cauchy slice of $M$ consisting of the fixed points of $T$. Let $a$ be a $T$-invariant wedge, i.e., $a=Ta$, or equivalently, $\eth a \subset \Sigma_T$. We define $e_T(a)$ as the wedge that satisfies
\begin{equation}
    a\subset e_T(a)~,~~~\eth e_T(a) \subset \Sigma_T~~~\mbox{and}~~~\tilde\partial a = \tilde\partial e_T(a)
\end{equation}
and which has the smallest generalized entropy among all such wedges~\cite{Bousso:2022hlz}.\footnote{Note that $e_T(a)$ defined here is the domain of dependence of the spatial region $E(a)=e_T(a)\cap \Sigma_T$ defined in \cite{Bousso:2022hlz}.}
\end{defn}

\begin{thm}
    With $a$ and $\Sigma_T$ as above,
    \begin{equation}
        e_T(a) = \emin(a) = \emax(a)~.
    \end{equation}
\end{thm}

\begin{proof}
    We first show that $e_T(a)\in F(a)\cap G(a)$. Property I and i listed in Defs.~\ref{emaxdef} and \ref{emindef} are trivially satisfied. Since $e_T(a)=Te_T(a)$, $e_T(a)$ must be normal or antinormal at every point $p\in\eth e_T(a)$. By arguments analogous to the proof of Lemma~\ref{thm:emaxextremal}, $e_T(a)$ is normal at $\eth a$ and extremal elsewhere. Hence $e_T(a)$ satisfies properties II and ii. Finally, it is easy to see that $\Sigma = e_T(a) \cap \Sigma_T$ and $\Sigma' = e_T'(a)\cap \Sigma_T$  satisfy properties III and iii, respectively.

    Since $e_T(a)\in F$, we have $e_T(a)\subset \emax(a)$. Similarly, since $e_T(a)\in G$, we have $\emin(a)\subset e_T(a)$. Hence $\emin(a)\subset e_T(a)\subset \emax(a)$. But Theorem~\ref{thm:emaxemin} established that $\emax(a)\subset \emin(a)$. Hence all three sets must be equal.
\end{proof}

\subsection{Asymptotic Bulk Regions}
\label{asymptotic}

\begin{defn}\label{emaxbdydef}
The \emph{max- and min-entanglement wedges of a conformal AdS boundary region} have been defined as follows~\cite{Akers:2020pmf,Akers:2023fqr}:\footnote{As usual, we replace min- and max-entropies by von Neumann entropies here even though for boundary regions $B$ this distinction is necessary to have $\maxEW(B) \neq \minEW(B)$.} Let $M$ be asymptotically Anti-de Sitter, and let $B$ be a partial Cauchy surface (a ``spatial region'') of the conformal boundary $\delta M$. Let $B'$ be the complement of $B$ on a (full) Cauchy surface of $\delta M$. Let $F(B)\equiv \set{f:\mathrm{I}\,\wedge\,\mathrm{II}\,\wedge\,\mathrm{III}}$ be the set of all wedges that satisfy the following properties:
\begin{enumerate}[1.]
\item $\tilde\eth f = B$\,;
\item $f$ is anti-normal;
\item $f$ admits a Cauchy slice $\Sigma_{\maxEW}$ such that for any wedge $h \neq f$ with $\tilde\eth h = B$ and $\eth h\subset \Sigma$, $\S(h)>\S(f)$.
\end{enumerate}
  The max-entanglement wedge $\maxEW(B)$ is their wedge union:
   \begin{equation}
       \maxEW(B) \equiv \Cup_{f\in F(B)}\, f~.
   \end{equation}
 Let $G(B)\equiv \set{g: \mathrm{i}\, \wedge\, \mathrm{ii} \,\wedge\, \mathrm{iii}}$ be the set of all wedges that satisfy the following properties: 
    \begin{enumerate}[i.]
        \item $\tilde\eth g = B$\,;
        \item $g$ is normal;
        \item $g'$ admits a Cauchy slice $\Sigma'_{\minEW}$ such that for any wedge $h \neq g$ with $\tilde\eth h = B$ and $\eth h\subset \Sigma'$, $\S(h)>\S(g)$.
    \end{enumerate} 
  The min-entanglement wedge $\minEW(B)$ is their intersection:
   \begin{equation}
       \minEW(B) \equiv \cap_{g\in G(B)}\, g~.
   \end{equation} 
\end{defn}

\begin{thm}\label{thm:emaxtomaxEW}
    Suppose that the wedge $a$ satisfies $a\subset \maxEW(\tilde\eth a)$, and that $\emax(a)$ is antinormal (and hence extremal by Theorem~\ref{thm:emaxextremal}).
    Then $\emax(a)=\maxEW(\tilde\eth a)$.
\end{thm}

\begin{proof}
    Any wedge $f \in F(\tilde\eth a)$ satisfies all the properties required of sets in $F(a)$ \emph{except} that $\Sigma_{\maxEW}$ need not satisfy IIIa. By footnote~\ref{asymmetric}, this suffices to apply the construction in the proof of Theorem~\ref{lem:emaxpropshare}, with $\Sigma_1=\smax(a)$ and $\Sigma_2=\Sigma_{\maxEW}$, to establish that $\emax(a)\Cup f\in F(a)$. This result is consistent with the definition of $\emax(a)$ only if $f \subset \emax(a)$. Hence $\maxEW(\tilde\eth a) \subset \emax(a)$.

    Conversely, with $\Sigma_1=\Sigma_{\maxEW}$ and $\Sigma_2=\smax(a)$, the same construction establishes that $\emax(a) \in F(\tilde\eth a)$. Hence $\emax(a)\subset\maxEW(\tilde\eth a)$. 
\end{proof}

\begin{cor}
The requirement that $\emax(a)$ is antinormal in Theorem \ref{thm:emaxtomaxEW} can be replaced by a requirement that $a$ is antinormal.
\end{cor}
\begin{proof}
    If $a$ is antinormal, it follows immediately that $\emax(a)$ is antinormal by property II and strong sub-additivity.
\end{proof}

\begin{thm}
    Suppose that the wedge $a$ satisfies $a\subset \minEW(\tilde\eth a)$. Then $\emin(a)=\minEW(\tilde\eth a)$.
\end{thm}

\begin{proof}
    The definitions of $G(a)$ and $G(\tilde\eth a)$ are identical except for the requirement that $a \subset g$ for all $g \in G(a)$. But, since we are told $a \subset \minEW(\tilde\eth a)$, this additional condition is already satisfied by all $g \in \minEW(\tilde\eth a)$, and so the two sets agree.
\end{proof}

\subsection{Examples}
\label{examples}

We will now analyze the examples shown in Fig.~\ref{fig:emin}. In the first example, the bulk input region $a$ is an asymptotic region in AdS whose outer boundary is the conformal boundary portion $A$, and whose inner boundary has negligible area because it is everywhere nearly null. (This can easily be arranged by ``wiggling'' the boundary of a static region up and down in time.) One finds that that $e_{\rm min}(a)=\EW($A$)$ and $e_{\rm max}(a)=a$.

The second example is spherically symmetric. The bulk input region extends into the interior of a two-sided black hole. In this case, one finds again that $e_{\rm max}(a)=a$. But $e_{\rm min}(a)$ extends further, to the horizon of the black hole. The edges of $e_{\rm min}(a)$ and $e_{\rm max}(a)$ are null separated; they are the boundaries of a lightsheet $L$~\cite{Bousso:1999xy}. We do not currently have an intuitive interpretation of this result.

\section{Discussion}
\label{discussion}

\subsection{Physical Interpretation}

The entanglement wedges $\emax(a)$ and $\emin(a)$ obey certain nontrivial properties. If the number of qubits that can flow through a surface $\gamma$ is given by $\A(\gamma)/(4G \log 2)$, all the information from $\emax(a)$ can flow into $a$, while all the information outside $\emin(a)$ can flow away from $a$. (More precisely, Theorems~\ref{lem:emaxpropshare} and \ref{lem:eminpropshare} establish that this quantum capacity is sufficient to achieve one-shot state merging respectively into and away from $a$.) Moreover, $\emin$ obeys nesting (Theorem~\ref{thm:nesting}); $\emax$ and $\emin$ obey a no-cloning relation (Theorem~\ref{thm:nocloning}); and when they concide, their generalized entropy obeys strong subadditivity (Theorem~\ref{thm:ssa}).

We now discuss the physical interpretation of $\emax(a)$ and $\emin(a)$ suggested by these properties. It will be important to compare and contrast this with the conventional interpretation of the entanglement wedges of a boundary region $B$, $\maxEW(B)$ and $\minEW(B)$, so we begin by reviewing the latter.

The meaning of $\maxEW(B)$ and $\minEW(B)$ is quite clear in the context of the AdS/CFT correspondence. The CFT is a nonperturbative completion of the semiclassical gravitational theory (or perturbative string theory) in the bulk. When we restrict to the semiclassical regime --- technically, by specifying a \emph{code subspace} in the CFT --- this becomes a duality between two equivalent descriptions. Viewed as a map from the bulk to the boundary, the duality is an isometry that implements a form of quantum error correction. It is possible, therefore, to define restrictions of the duality that relate a subregion $B$ of the boundary to an appropriate subset of the bulk. After fixing the code subspace, the wedge $\maxEW(B)$ characterizes the region that can be fully reconstructed from $B$; and $\minEW(B)$ is the smallest bulk region outside which no information can be reconstructed. 

This interpretation is supported by several nontrivial properties that $\maxEW(B)$ and $\minEW(B)$ have been shown to obey. Historically, the initial evidence was not directly related to reconstruction, but to the related~\cite{Harlow:2016vwg,Akers:2021fut} entropy formula, Eq.~\eqref{eq-rt}:\footnote{In fact, subregions in a relativistic field theory have a type-III von Neumann algebra, in which an entropy is not defined. We may sidestep this subtlety here by putting the CFT on a lattice with fine enough spacing.} if $\maxEW(B)=\minEW(B)$, then $S(B) = \S[\EW(B)]$. This can be verified explicitly when $S(B)$ can be computed in the CFT. Moreover, the entanglement wedges of disjoint boundary regions obey strong subadditivity of the \emph{generalized} entropy, which suggests that $\S[\EW(A)],\S[\EW(B)],\S[\EW(C)]$ really correspond to the ordinary von Neumann entropies of subsystems $A,B,C$ of a quantum mechanical system (the CFT) \cite{Headrick:2007km, Wall:2012uf}. 

Consistent with their information theoretic interpretation, the wedges $\maxEW(B)$ and $\minEW(B)$ also obey nesting, as well as no-cloning, which follows immediately from complementarity: $\maxEW(\bar B)'=\minEW(B)\supset\maxEW(B)$. These properties were in fact some of the primary early evidence that entanglement wedges were more than just a way of computing CFT entropies, and in fact described the bulk dual of the boundary region $B$ \cite{Czech:2012bh,Wall:2012uf}. 

On the other hand, no duality analogous to AdS/CFT is available in the general setting in which $\emax(a)$ and $\emin(a)$ are defined. There is no manifest, complete quantum mechanical system such as the CFT, to whose subsystems these entanglement wedges could be associated. On the ``bulk side,'' $\emax(a)$ and $\emin(a)$ admit a description in terms of semiclassical gravity, exactly like $\maxEW(B)$ and $\minEW(B)$. But the analogue of the ``boundary side,'' $a$, is now also a gravitating region, whereas $B$ was a purely quantum mechanical system. In particular, it is obvious that the description of $a$ in terms of semiclassical gravity --- its only currently known description --- cannot be equivalent by a duality to that of $\emax(a)$. 

Thus, we are not yet in a position to interpret $a$ and its entanglement wedges in terms of a duality, i.e., as two known, equivalent representations of the same data. Instead, the existence and properties of $\emax(a)$ and $\emin(a)$ suggest that $a$ should possess an \emph{unknown}, new structure that is purely quantum-mechanical (and thus distinct from its semiclassical description), in which all information in $\emax(a)$, and none outside $\emin(a)$, can be represented. We will denote this unknown quantum mechanical system by $\ba$.

Our viewpoint, then, is that the relation between $\ba$ and the entanglement wedge of $a$ should be a new  holographic correspondence in general spacetimes whose details we have yet to learn. Fortunately, we can already infer some aspects of the unknown system $\ba$ from the properties of $\emax(a)$ and $\emin(a)$, as follows.

By its definition, information in $\emax(a)$ can be transmitted towards $a$, across any intermediate homology surface between their edges, with resources set by the area of that surface. Information outside $\emin(a)$ can similarly be transmitted away from $\emin(a)$. Exactly the same properties hold for $\maxEW(B)$ and $\minEW(B)$ with respect to $B$, where they reflect the fact that $B$ encodes all information in $\maxEW(B)$ and none outside $\minEW(B)$. This analogy suggests that $\ba$ encodes all information in $\emax(a)$ and none outside $\emin(a)$.

The purely \emph{quantum-mechanical} nature of $\ba$ is suggested by Theorem~\ref{thm:ssa}, which states that $\S[e(a)],\S[e(b)],\S[e(c)]$ obey strong subadditivity as though they were von Neumann entropies. This feature distinguishes $\ba$ from $e(a)$ and prevents us from trivially defining the former as the latter. For $\ba$ to possess a von Neumann entropy, it must be manifestly a quantum mechanical subalgebra; thus it cannot be defined as a spacetime region in semiclassical gravity.

The nesting of $\emin$, Theorem~\ref{thm:nesting}, suggests that $\ba$ should be constructed from resources that can be associated to $a$, even if they transcend the semiclassical description of $a$. If $\emin(a)$ is the smallest region outside which nothing can be reconstructed from $\ba$, then its growth as $a$ is increased can only be explained by such a relation.

The failure of $\emax$ to obey nesting (which has no analogue for $\maxEW$) suggests that one of the resources that determine $\ba$ is the amount of entanglement between $a$ and $\emax(a)\cap a'$. In fact, the example of Hawking radiation after the Page time suggests that the additional information that distinguishes $\ba$ from $a$ can somehow be made to appear precisely in those physical degrees of freedom in $a$ that are entangled with $\emax(a)\cap a'$ in the semiclassical description. (In most other examples, these would be dominated by short-wavelength degrees of freedom near the boundary of $a$, most of them near the Planck scale; and it is not clear how the information can be caused to appear there. We will return to this question when we discuss summoning, below.)

The failure of $\emax$ to obey nesting also implies that the interior of $a$ in itself is not simply related to the resources determining $\ba$. To see this, consider a wedge $a$ that is a proper subset of $\emax(a)$. Then $a$ can be enlarged by deforming its edge infinitesimally into $\emax(a)$ and wiggling the corresponding portion so that the area decreases significantly. This will decrease the entanglement resource and hence can decrease $\emax$, even though it will have increased the interior of $a$.

Finally, and perhaps most importantly, the no-cloning Theorem~\ref{thm:nocloning} suggests that an observer with access to $a$ can somehow summon information from the spacelike related region $\emax(a)\cap a'$. The cloning of a quantum state is definitely inconsistent only if it can actually be verified by an observer. If $\ba$ and $\bb$ were merely formal representations of the quantum information in their entanglement wedges, then there would be no observable paradox if those wedges failed to be disjoint. The fact that the theorem forbids an overlap suggests that $\ba$ and $\bb$ can in principle be operationally accessed by a single observer (or a pair of observers who can subsequently meet), without completely destroying the entire spacetime. If summoning is possible, then the theorem has an essential role: it prevents the simultaneous summoning of information in the overlap of $\emax(a)\cap \emin(b)$ to two distant observers with access respectively to $a$ and to $b$, who could otherwise later verify that they have cloned an unknown quantum state.

\subsection{Summoning of Spacelike Related Information}

The properties of $\emax$ and $\emin$ have led us to the conjecture that a bulk observer in $a$ can summon spacelike-related information from $\emax(a)\cap a'$. This is a radical and novel proposition. Moreover, if information can be summoned, then presumably this action can be reversed, perhaps after applying a local operator to the quantum information while it is controlled by the bulk observer in $a$. In this manner, any simple operator in $\emax(a)\cap a'$ can be enacted from $a$. Thus, information can be manipulated at spacelike separation from $a$.

We do not know by what protocol a bulk observer might accomplish the summoning task. But spacelike information transfer is obviously inconsistent with locality. Thus, summoning must involve a breakdown of the semiclassical description of an appropriate portion of the bulk --- perhaps $\emax(a)\cap a'$ --- so that locality cannot be invoked against the process. 

To gain further intuition, let us retreat for a moment to more familiar ground of entanglement wedges of AdS boundary regions and of external quantum systems. We will see that the semiclassical bulk description does break down in AdS/CFT, when a CFT observer instantaneously summons information from deep in the bulk. But first, consider Hawking radiation extracted from a gravitating spacetime $M$ and stored in an external bath $R$.

After the Page time, $EW(R)$ contains an island $I$ inside the black hole. Assuming unitarity, we know that the information that in the semiclassical description appears to be in the island is actually encoded in the quantum system $R$. At most, ``summoning'' information from the island (for example, the state of the star that collapsed and formed the black hole) only requires decoding the information in $R$ into an easily accessible form.

If the same radiation resides instead in an asymptotic region $a\subset M$, then $e(a)\subset I$. We do not expect any significant difference to the previous case: the information in $I$ can be decoded from the radiation in $a$. Although gravity is present in $a$, it is irrelevant, as this task involves only low-energy quantum field theory degrees of freedom contained within $a$. The extent to which information must be summoned is limited to the decoding task, and decoding is a completely semiclassical process within $a$ itself.

When the Petz recovery map is implemented through a gravitational path integral~\cite{Penington:2019kki}, the act of decoding the radiation appears to ``move'' information from the island into $a$ or $R$ through a semiclassical Euclidean wormhole created by the process. But from the Lorentzian point of view, this process is still acausal, so the semiclassical description of $M$ as a whole breaks down in any case. With the above interpretation, it breaks down \emph{because} of decoding; otherwise, it breaks down earlier, at the Page time, when information seemingly in $I$ first begins to be available for decoding in $R$ or $a$.

For decoding the Hawking radiation in $a$, we have seen that it is sufficient to regard $\ba$ as the the quantum algebra generated by operators acting on low-energy quantum fields in $a$. Next we would like discuss summoning more generally. As a relatively simple example, consider the task of instantaneously summoning information from deep inside AdS into an asymptotic region $a$, say the exterior of a large sphere. In this case, involving only low-energy semiclassical quantum fields in $a$ appears to be insufficient. To understand this, let us begin by reviewing a related but better understood problem: how a boundary (or CFT) observer can summon information from deep inside the bulk.

\begin{figure}[t]
\begin{subfigure}{.3\textwidth}
  \centering
 \includegraphics[width = 0.94\linewidth]{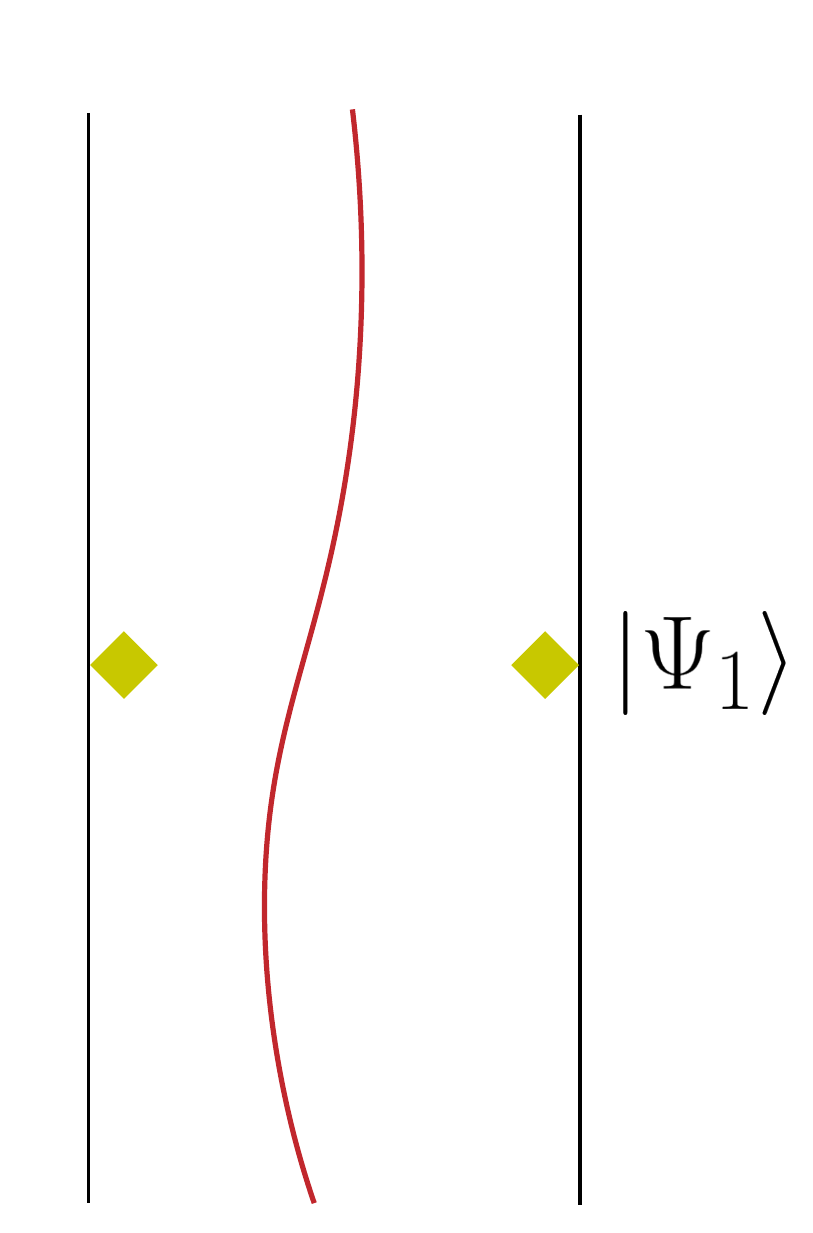}
\end{subfigure}
\begin{subfigure}{.3\textwidth}
  \centering
 \includegraphics[width = 0.94\linewidth]{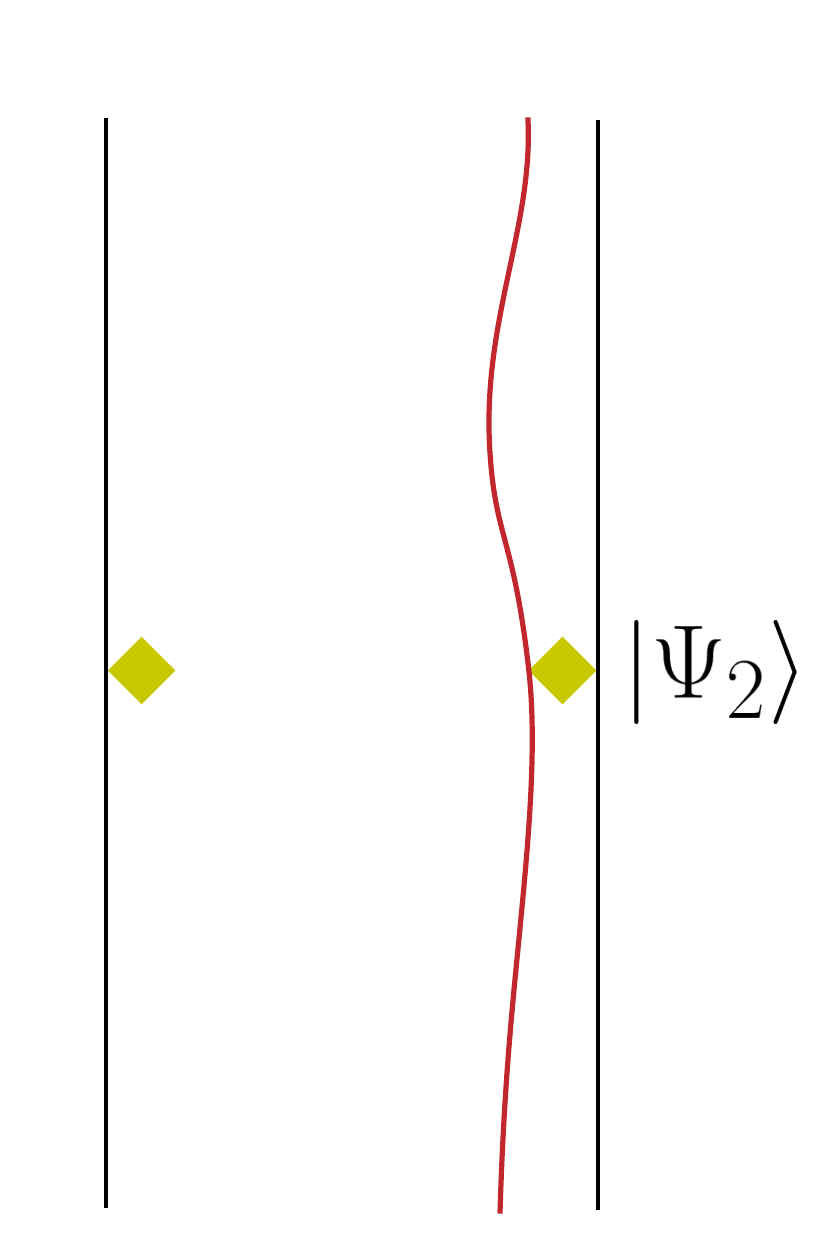}
\end{subfigure}
\begin{subfigure}{.35\textwidth}
  \centering
 \includegraphics[width = 0.8\linewidth]{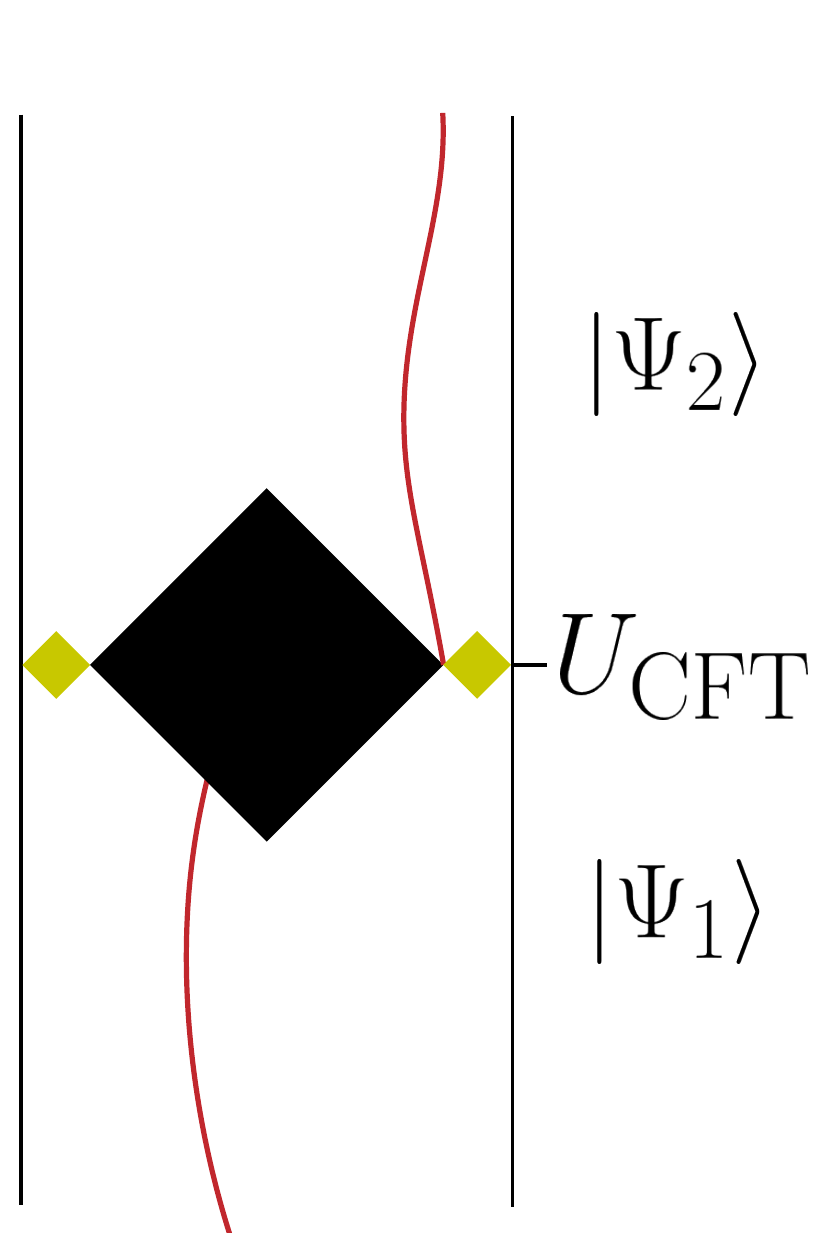}
\end{subfigure}
\caption{\emph{Left:} In the spacetime dual to the CFT state $\ket{\Psi_1}$, a qubit is located near the center of the bulk. \emph{Middle:} In the spacetime dual to the CFT state $\ket{\Psi_2}$, the same qubit is located at the edge of the asymptotic region $a$ (yellow diamond) around the time $t=0$. \emph{Right:} Consider the CFT in the state $\ket{\Psi_1}$ for $t<0$ and $\ket{\Psi_2}$ for $t>0$; this is implemented by acting with $U_{\rm CFT}$ at $t=0$. This boundary state admits no semiclassical bulk dual in the black patch, spacelike to $a$. Our results suggest that an asymptotic bulk observer in $a$ may be able to trigger this process.}
\label{fig:summon}
\end{figure}

Consider a CFT in the state $\ket{\Psi_1}$ at $t=0$, corresponding to a spacetime $M$ that is nearly empty AdS, with only one qubit in the unknown state $\ket{\phi}$ at its center. Let $a\subset M$ be the spatial exterior of a large sphere in the Wheeler-de Witt patch of the boundary time $t=0$ (see Fig.~\ref{fig:summon}). 

According to the AdS/CFT duality, the state $\ket{\psi}$ is a logical qubit encoded in the boundary CFT. Therefore, a ``CFT observer'' with full access to the CFT operator algebra (or indeed, with access to somewhat more than half of the boundary) could implement a unitary operator at $t=0$ that distills $\ket{\psi}$ into a localized CFT excitation:
\begin{equation}
    \ket{\Psi_2} = U_{\rm CFT} \ket{\Psi_1}~.
\end{equation}
We can arrange for this new state $\ket{\Psi_2}$ to have the property that its bulk dual is nearly empty AdS with a particle in the state $\ket{\phi}$ just outside the large sphere $\eth a$.

The CFT evolution we have just described obviously has no bulk dual consistent with semiclassical gravity. A particle cannot move in a spacelike way. Moreover, it is not clear at what bulk time this jump should be thought of as taking place. We can only say that in the causal past of the boundary time $t=0$, the bulk has a particle in the center; and in the causal future of the same boundary slice, the bulk has a particle in the asymptotic region $a$. But the Wheeler-de Witt patch of the slice $t=0$ has no semiclassical bulk interpretation; see Fig.~\ref{fig:summon}. 

Arguably, the process is somewhat less violent: $U_{\rm CFT}$ destroys a large portion of the Wheeler-de Witt patch, but it need not destroy all of it. Let $\lambda$ be small length scale on the boundary, such that the entanglement wedge of a boundary region of size $\lambda$ is much more shallow than $a$. Putting the CFT on a $\lambda$-spaced lattice should not alter our discussion, which suggests that $U_{\rm CFT}$ need not involve any such short wavelength modes. If we arrange for the particle to appear at the inner boundary of $a$, then locality does not forbid, and the above argument suggests, that the bulk evolution remains semiclassical in the asymptotic region $a$ throughout the process, preserving its geometric description. 

This allows us to view the above process as a blueprint for summoning information from $\emax(a)\cap a'$ into $a$. From the bulk point of view, the operator $U_{\rm CFT}$ has retrieved spacelike-related information from deep in the bulk, and placed it at the edge of the asymptotic region $a$. The process preserves the semiclassical description of $a$ but not of $\emax(a)\cap a'$ (which in this simple example is simply $a'$).

The existence of the boundary operator $U_{\rm CFT}$ does not tell us how an asymptotic observer in $a$ would accomplish the same task on their own initiative. But it may be possible. The full CFT algebra is generated by local CFT operators (modulo details about gauge constraints). By the extrapolate dictionary, a local CFT operator is dual to a quasi-local bulk operator near the asymptotic boundary. Hence, $U_{\rm CFT}$ should be contained in the algebra $\ba$ generated by quasi-local bulk operators in $a$. To be clear: these quasi-local bulk operators do not in general act only on light bulk quantum fields; generically the bulk dual of a heavy CFT operator will create a black hole. However, that black hole will still be localised near the asymptotic boundary, and hence the bulk operator is still contained in the algebra $\ba$.

In fact, an argument by Marolf \cite{marolf2009unitarity} suggests that $U_{\rm CFT}$ is contained in an algebra generated by only \emph{light} bulk QFT operators in $a$ along with the ADM Hamiltonian $H$. Any simple bulk operator at $t=0$ can be rewritten as an integral over bulk operators $\Phi(t)$ near the asymptotic boundary at times $t \in [-\pi/2,\pi/2]$~\cite{Hamilton:2006az}. In a quantum theory, we must therefore have $\Phi(t) = \exp(i H t ) \Phi(0) \exp(-i H t )$.

It follows that any simple operator in $a'$ is contained in the quantum algebra generated by $H$ along with light bulk quantum fields in $a$. It is crucial here that this the full algebra in \emph{nonperturbative} quantum gravity; in the classical limit the time evolution of asymptotic boundary operators is in general not analytic and hence is not determined by local data at the boundary \cite{Jacobson:2019gnm}. Of course, in principle, the ADM Hamiltonian $H$ can be determined solely by measuring the metric in $a$ with sufficient precision. However, that precision scales as $O(G)$ in the semiclassical limit. This means that the semiclassical graviton field $h_{\mu \nu} = \sqrt{G}\, \delta g_{\mu\nu}$ does not know about $O(1)$ fluctuations in energy; the ADM Hamiltonian is not a semiclassical bulk QFT operator. In contrast to the situation with Hawking radiation above, the algebra $\ba$ needs to contain more than just semiclassical bulk QFT operators if it is to encode the entanglement wedge $e(a)$. Any bulk observer who wants to summon information from deep in the bulk needs access to non-semiclassical -- presumably Planckian -- degrees of freedom.\footnote{An independent argument to this effect based on tensor network toy models was given in \cite{Bousso:2022hlz}.}

\subsection{Independence of Semiclassical Bulk Regions}

Seemingly independent regions in semiclassical gravity may in fact be the same fundamental degrees of freedom, in different guises. A classic example is the black hole interior and the Hawking radiation. Arguably, these are complementary descriptions of the same quantum system, i.e., different ways of representing the same quantum information~\cite{Susskind:1993if}. But given two bulk regions in an arbitrary semiclassical spacetime, no general method was known to determine whether they are truly independent, or partly or wholly complementary. 

In AdS/CFT, the fundamental degrees of freedom are known. At the fundamental level, independence reduces to the trivial condition that two boundary regions $A$ and $B$ are disjoint. This is equivalent to the condition that $\EW(A)$ and $\EW(B)$ are disjoint. It leads to a reasonable conjecture for a sufficient condition for the independence of gravitating regions in AdS: two bulk regions $a,b$ are independent if there exist disjoint boundary regions $A,B$ such that $a\subset \EW(A)$ and $b\subset \EW(B)$.

Bulk entanglement wedges allow us generalize this to a necessary and sufficient criterion in arbitrary spacetimes: two bulk regions $a,b$ are independent iff $a\subset e(b)'$ and $b\subset e(a)'$.

So far, we have neglected the difference between max- and min-entanglement wedges. This leads to a subtlety already in AdS/CFT, where $\minEW(B)$ and $\minEW(\bar B)$ can overlap despite $B$ and $\bar B$ being manifestly independent. Based merely on the characterization of $\minEW(B)$ as the smallest region about whose exterior $B$ has no information, this overlap makes it hard to rule out the possibility of cloning. Additional structure in the map from bulk to boundary must prevent that. On the other hand, if $\maxEW(B)$ and $\maxEW(\bar B)$ overlapped there would definitively be a cloning paradox: information that passed through both $\maxEW(B)$ and $\maxEW(\bar B)$ would necessarily be reconstructible on both $B$ and $\bar B$. The true condition that $\maxEW(B)$ cannot overlap with $\minEW(\bar B)$ (and vice versa) is therefore both somewhat stronger than the minimal condition necessary for the theory to avoid a provable cloning paradox, but weaker than needed to prove (from the bulk alone) that no such paradox exists. To understand why this particular condition is true (but not anything stronger), we have to remember that $B$ and $\bar B$ are not merely independent subsystems, but in fact complementary subsystems. As a result, all information not encoded in $B$ must be encoded in $\bar B$ and vice versa; consequently we always have $\minEW(\bar B) = \maxEW(B)'$.

A similar ambiguity exists regarding the condition for two bulk wedges $a, b$ to be independent. It is reasonable to expect that $a$ and $b$ should be totally independent -- and hence any cloning of information in both $a$ and $b$ should be definitively paradoxical -- only if $a\subset \emin(b)'$ and $b\subset \emin(a)'$. As above, however, the provable result is somewhat stronger than the minimal condition necessary to avoid a definitive paradox. Instead, we find that our no-cloning condition holds whenever either $a\subset \emin(b)$ and  $b\subset \emax(a)$, or $a\subset \emax(b)$ and  $b\subset \emin(a)$. Unlike in the discussion of overlaps of boundary entanglement wedges, we do not know a deeper principle -- analogous to $B$ and $\bar B$ being complementary subsystems -- that picks out this particular condition.

\subsection*{Acknowledgements}
This work was supported in part by the Berkeley Center for Theoretical Physics; by the Department of Energy, Office of Science, Office of High Energy Physics under QuantISED Award DE-SC0019380 and under contract DE-AC02-05CH11231. RB was supported by the National Science Foundation under Award Number 2112880. GP was supported by the Department of Energy through an Early Career Award; by the Simons Foundation through the ``It from Qubit'' program; by AFOSR award FA9550-22-1-0098; and by an IBM Einstein Fellowship at the Institute for Advanced Study.

\bibliographystyle{JHEP}
\bibliography{covariant}

\end{document}